\documentclass{JHEP3}
\usepackage{epsfig}

\newcommand{\be}{\begin{equation}}
\newcommand{\ee}{\end{equation}}
\newcommand{\bea}{\begin{eqnarray}}
\newcommand{\eea}{\end{eqnarray}}
\newcommand{\IR}{\mathbb{R}} 

\newcommand{\IQ}{\mathbb{Q}}
\def\IZ{\relax\ifmmode\hbox{Z\kern-.4em Z}\else{Z\kern-.4em Z}\fi}

\newcommand{\non}{\nonumber \\}

\def\half{{1 \over 2}} 

\def\del{{\partial}}

  \def\eps{\epsilon}

\newcommand{\sbsection}[1]{\vspace{.5cm} \noindent {\it #1}}

\def\scri{{\cal I}}
\def\phiom{\phi_\omega}
\def\RN{Reissner-Nordstr\"{o}m }
\def\Schw{Schwarzschild }
\def\KS{Kruskal-Szekeres }

\def\({\left(}
\def\){\right)}

\def\br{\left[}
\def\kt{\right]}

\def\exp{{\rm exp}}

\def\ch{{\rm cosh}}

\preprint{{\tt hep-th/0401209}}
\title{On the Resolution of the Time-Like Singularities in
Reissner-Nordstr\"{o}m and Negative-Mass Schwarzschild}

\author{Amit Giveon\footnotemark[1], Barak Kol\footnotemark[1],
 Amos Ori\footnotemark[2], Amit Sever\footnotemark[1] \\

\footnotemark[1]
 Racah Institute of Physics \\
 The Hebrew University \\
 Jerusalem 91904 \\
 Israel\\
{\tt giveon, barak\_kol, asever @phys.huji.ac.il}
\\

\footnotemark[2] Department of Physics\\
  Technion-Israel Institute of Technology\\
  Haifa 32000\\
  Israel \\
  \email{amos@physics.technion.ac.il} }

\abstract{Certain time-like singularities are shown to be resolved
already in classical General Relativity once one passes from
particle probes to scalar waves. The time evolution can be defined
uniquely and some general conditions for that are formulated. The
\RN singularity allows for communication through the singularity
and can be termed ``beam splitter'' since the transmission
probability of a suitably prepared high energy wave packet is
25\%. The high frequency dependence of the cross section is
$\omega^{-4/3}$. However, smooth geometries arbitrarily close to
the singular one require a finite amount of negative energy
matter. The negative-mass \Schw has a qualitatively different
resolution interpreted to be fully reflecting. These 4d results
are similar to the 2d black hole and are generalized to an
arbitrary dimension $d>4$. }

\begin{document}

\section{Introduction}

The importance of singularities in General Relativity and their
resolution is well appreciated. The most intriguing ones are
considered to be those inside black holes, and the Big-Bang-like
cosmological singularities. String theory offers some cases where
a time-like singularity gets resolved: the orbifold, the flop, and
the conifold. These are all singularities in compact factors of
spacetime, the last two occurring in Calabi-Yau manifolds, and
their resolution often involves adding light matter (``twisted
sectors,'' ``wrapped D-branes'') which lives on the singularity
(for a review, see \cite{Polchinski} and references therein).
 Recently, there was a renewed effort to include
time-dependence in string theory (for a review, see
\cite{GRS-review} and references therein). Ultimately, one would
like to understand space-like singularities such as Schwarzschild
and cosmological singularities, but it would be fair to say that
there were no breakthroughs yet.

The study described here was motivated by bold attempts to cross
the singularity in a stringy black-hole model. The 2d black hole
is a case where an algebraic coset construction requires one to
add to spacetime additional manifolds which are glued together
over the singularity (see figures
\ref{PenroseRN},\ref{PenroseSchw}). For the uncharged 2d black
hole it was found \cite{dvv} that waves sent from the
``additional'' infinity are fully reflected, while a charged 2d
black hole does have a non-zero transmission amplitude across the
singularity and, moreover, wave functions are smooth there
\cite{GiveonRabinoviciSever}.

Here we show that there is nothing intrinsically stringy in those
calculations~\footnote{There are $\alpha'$ corrections in string
theory, which were computed in the 2d case. These give rise to
extra phases in scattering amplitudes which vanish in the
semi-classical limit.}. Actually this interesting phenomenon
happens once one considers the wave equation on this background,
namely, by passing from point-particle probes~\footnote{The usual
definition of a singular spacetime is ``geodesic-incompleteness,''
namely, incomplete time-evolution for particle probes.} to waves.
More precisely, even though the wave equation is singular, the
ordinary differential equation (ODE) obtained after separating
time, and whose solutions we denote by $\phiom$, allows for a
unique, smooth continuation through the singularity.
Interestingly, this property is not special to the 2d charged
black hole, but it generalizes to the 4d charged
Reissner-Nordstr\"{o}m (RN) black hole, as well as to $d>4$ RN
black-holes (moreover, for the 2d and 4d cases the relevant ODE is
actually completely regular). This observation strongly suggests
that the above classical RN spacetimes should be viewed as being
made of two parts which are naturally connected across the
singularity at $r=0$. The two parts are
 (i) $r>0$ which is a usual (positive-mass) RN black
hole, and
 (ii) $r<0$, the region beyond the singularity,
which is a negative-mass spacetime.

Another motivation for considering the negative-$r$ part of RN
comes from the realistic astrophysical black holes, which are
known to be rotating in general \cite{bardeen,thorne,central}. In
the analytically-extended Kerr geometry (describing a stationary
spinning black hole), the Boyer-Lindquist coordinate $r$ goes
smoothly from positive to negative values. The region $r<0$ is an
asymptotically-flat universe (distinct from the original
positive-$r$ external universe). There is a curvature singularity
at $r=0$, but this singularity is a ring rather than a
hypersurface. A timelike geodesic heading towards $r=0$ will
generically avoid the ring singularity, and smoothly pass
``through the ring'' into the negative-$r$ asymptotic region.
Obviously, wave packets will also pass through the ring, though
with partial reflection. Spherically-symmetric charged black holes
were often considered as useful toy models for the more realistic
spinning black holes, due to the remarkable similarity in the
inner structure of the two black-hole types. From this perspective
we may regard the RN solution analyzed here as such a simplified
toy model for a spinning black hole.

Initially one would expect the time evolution of a wave to be
ill-defined in the presence of a singularity since some boundary
condition (b.c.) needs to be supplied there. Given the smooth and,
in particular, univalued nature of the eigen-functions $\phiom(r)$
there is a natural way to formulate a unique time evolution simply
by following the usual recipe: decompose any incoming wave packet
into $\phiom$ and then endow each component with an $\exp(i\,
\omega\, t)$ time dependence. Since the eigen-functions
$\phiom(r)$ are smooth at $r=0$, it is just natural in this
formulation to consider both sides of the singularity to be
connected and communicating.

For negative-mass \Schw the situation is different from the
charged case, though our approach still provides a unique time
evolution. In this case the $r=0$ singularity ``flips'' from being
time-like at $r<0$ to space-like at $r>0$, suggesting that gluing
these two pieces would make little physical sense. Analyzing the
wave equation suggests that indeed in the Schwarzschild case the
field in the $r<0$ region evolves without any connection to the
$r>0$ region. Unlike the RN case, here the eigen-functions
$\phiom(r)$ generically have a log singularity, hence they cannot
be extended across the singularity in a univalued fashion. Yet, a
single regular $\phi_{1\omega}(r)$ solution exists for each
$\omega$. In this case a natural choice of boundary condition
suggests itself -- the so-called ``regularity b.c.'' -- i.e. the
demand that $\phi_{1\omega}(r)$ be bounded at the singularity.
Since we have a single regular solution for each $\omega$, any
wave packet coming from infinity will have a unique decomposition
into these regular functions $\phi_{1\omega}(r)$, and hence a
unique time evolution. (This is to be contrasted with the RN case,
in which there are two regular solutions for each $\omega$, and
therefore, the time evolution depends on initial data from both
sides of the singularity.) This ``regularity b.c.'' can be
physically interpreted as a reflecting b.c. While normally one can
define reflecting b.c. either by Dirichlet or by Neumann b.c. (or
a mix) the ``regularity b.c.'' at Schwarzschild's $r=0$ do not
leave us such a choice~\footnote{It was found already in
\cite{IshibashiHosoya} by different methods that there exists a
unique time evolution for negative-mass Schwarzschild, which is
probably the same one we explicitly describe and physically
interpret here.}.

Thus we find {\it a natural way to define (scalar wave) physics in
the presence of certain  time-like singularities} (by ``defining
physics'' we mean a recipe that allows to predict the time
evolution in spacetime) simply by passing from particle probes to
waves (we call such singularities ``wave-regular''). We discuss
two mechanisms: for RN we continue the space-time across the
singularity allowing for cross-communication, while for
negative-mass \Schw the ``regularity b.c.'' are unique and define
complete reflection.

We would like to pause to make a few comments on this picture:
\begin{itemize}

\item We consider this to be an important observation in the
search for a resolution of black hole singularities, at least
those of the time-like type. Yet, for this observation to be more
than a mere curiosity it is required that several additional tests
be satisfied such as regularity of higher-spin waves and
regularity against various perturbations to the equations (some of
which we perform).

\item At the classical level there are issues of interpretation,
for instance, the existence of these spacetimes or others with the
relevant singularities, the observers living in them and
implications to Cosmic Censorship, which we discuss in section
\ref{discussion}.

\item At the level of quantum gravity additional issues arise,
including  the validity of the wave equation and Hawking radiation
whose discussion we defer to section \ref{discussion} as well.
Here we only want to stress that one should be cautious not to
jump to conclusions and ``legitimize'' spacetimes with such
singularities as possible solitons.

\end{itemize}

After reviewing the RN and \Schw backgrounds in section
\ref{ReviewRN}, we explain our main observation in section
\ref{smooth-section}. We formulate the wave evolution for
arbitrary wave-packets coming from infinity (i.e. from large
negative $r$). However, the issue of specifying initial conditions
on a spacelike hypersurface crossing the $r=0$ singularity becomes
non-trivial and is relegated to section \ref{conditions-section}.

Once the unique evolution of the field is formulated, we are in a
position to investigate scattering phenomena. In section
\ref{cross-section-section} we study the physical properties of
the RN singularity by computing the cross section for
transmission. Even though the ODE for $\phiom$ is regular, the
effective potential diverges as $V_{\rm eff} \sim -c/r^{*~2}$ in
the vicinity of the singularity, where $r^*=r^*(r)$ is the
canonical ``tortoise'' coordinate. Therefore, one would expect the
singularity to have an effect even on  incoming waves of very high
energy -- unlike a regular point in spacetime, which becomes
``transparent'' in this limit. Indeed we show that the high-energy
transmission amplitude for fixed angular momentum $l$ approaches
25\% rather than 100\%. In this respect the singularity behaves as
a beam-splitter. Summing over $l$ we find the high-energy
dependence of the total cross section to be $\propto
\omega^{-4/3}$ where $\omega$ is the frequency of the incoming
plane wave. This behavior stands between the $\omega^0$ dependence
of absorption for a black hole with well-defined area, and the
$\omega^{-2}$ dependence of scattering off an elementary particle
in field theory.

Generalizations of the backgrounds to higher dimensions and for
the 2d black hole are presented in section \ref{generalizations}.
We determine the transmission amplitude and total cross-section as
in the 4d case. The only significant difference is that while in
4d (and 2d) RN the variable-separated wave equation is regular,
for $d>4$ it is singular-regular. Yet, all the solutions
$\phiom(r)$ are still univalued for a continuation through the
singularity. For Schwarzschild, on the other hand, there are no
significant changes in passing from $d=4$ (and $d=2$) to $d>4$.
Hence, we also present the results for $d=4$ Schwarzschild in
section 5.

Next we challenge our picture by adding various perturbations in
section \ref{perturbations-section}. So far we discussed only
probes which were minimally-coupled scalar fields. Here we preform
several perturbations: first we consider a field with both mass
and charge, then we add interactions, and finally we consider the
effect of back-reaction.

In section \ref{conditions-section} we discuss some of the
conditions for a general timelike singularity to be wave-regular,
including the necessary constraint on initial conditions specified
on a spacelike hypersurface crossing the singularity, the
uniqueness of decomposition of an incoming wave-packet, and the
univalued property of the eigen-functions $\phiom$. In section
\ref{discussion} we summarize and discuss the issues mentioned
above, and list some open questions. Some technical details are
given in the appendices.

Note that, in principle, this study could have been carried out a
long time ago and, indeed, it turns out that there were some
attempts in this direction. However, it seems that none took the
conceptual freedom to cross the singularity and that actually led
several authors to pronounce RN to be wave-singular. Wald
\cite{Wald1980} gave a general discussion on extending the
definition of the wave equation to include non globally-hyperbolic
spacetimes relying on the mathematical notion of the ``Friedrichs
extension.'' Horowitz and Marolf \cite{HorowitzMarolf} showed that
some dilatonic black holes are wave-regular, where they consider a
Schr\"{o}dinger-like equation rather than the wave equation and
demonstrate ``essentially self-adjointness'' of the associated
operator. Ishibashi and Hosoya \cite{IshibashiHosoya} were able to
claim wave-regularity for negative-mass \Schw by going back to the
wave equation together with changing the metric in function space
accordingly from the ordinary $L^2$-metric to the Sobolev metric,
but using the same criterion of ``essentially self-adjointness.''
However, they do not comment on the physical nature of this
resolution. Related work also appears in \cite{Peeters:1994jz}.
Finally, Jacobson \cite{Jacobson} observed in passing the
regularity of the time-separated wave equation near the RN
singularity while studying the semiclassical decay of
near-extremal black holes.

{\bf Note added at second version:} it is interesting to consider
smooth approximations to the singular geometries. We found that if
one smoothly approximates the RN singularity one is required to
add a finite amount of negative energy matter, which is what one
would expect due to the wormhole nature of the singularity region.
The details of the calculation are given in section \ref{ReviewRN}
in the part titled ``Negative-energy content of the smeared
singularity.'' The subsequent ``perspectives on the $r<0$ region''
part in section \ref{ReviewRN} as well as the discussion section
should be read with this additional fact in mind. We gratefully
acknowledge the valuable comments of G. Gibbons  and J. Katz on
this matter. See \cite{Lynden-BellKatz,Lynden-BellKatz2} for a
closely related case.

\section{Review of the Reissner-Nordstr\"{o}m and Schwarzschild backgrounds}
\label{ReviewRN}

\vspace{0.5cm} \noindent {\it Basic features of the
Reissner-Nordstr\"{o}m geometry}

The background we wish to consider is the Reissner-Nordstr\"{o}m
(RN) geometry \cite{Reissner,Nordstrom}. In Schwarzschild
coordinates $(t,r,\theta ,\varphi )$ the metric is
\begin{equation}
ds^{2}=-f\,dt^{2}+f^{-1}\,dr^{2}+r^{2}\,d\Omega ^{2},
\label{RNmetric}
\end{equation}
where
\[
f=1-{\frac{2\,M}{r}}+{\frac{Q^{2}}{r^{2}}}
\]
and $d\Omega ^{2}$ is the two-sphere metric,
\[
d\Omega ^{2}=d\theta ^{2}+\sin ^{2}\theta \,d\varphi ^{2}.
\]
The electromagnetic four-potential is given by \footnote{The minus
sign was entered in order to conform with the rest of our sign
conventions and to produce (\ref{Veff-geod}) correctly.}
\[
A=-{\frac{Q}{r}}\,dt ~.
\]
For $Q=0$ the metric reduces to the Schwarzschild metric
\cite{Schw}.

For the (under-extremal) non-extremal black-hole case, $M>|Q|>0$
$f$ has two positive roots, located at
\[
r_{\pm }=M\pm \sqrt{M^{2}-Q^{2}}.
\]
Both $r=r_{+}$ and $r=r_{-}$ are null hypersurfaces, known as the
event horizon and the inner horizon, respectively (in
Schwarzschild there is only one such root, $r=r_{0}:=2\,M$, which
is the event horizon). Note that the roles of $r$ and $t$ as
space-like and time-like coordinates interchange at each root of
$f$.

The line element (\ref{RNmetric}) is singular at the horizons, but
this is merely a coordinate singularity which may be removed by
transforming to other coordinates (e.g. the Kruskal-Szekeres
coordinates discussed below). On the other hand, $r=0$ is a true
physical singularity (e.g. the Ricci scalar diverges in the RN
case), and we focus on this singularity in the present paper.

\begin{figure}
\centerline{\epsfxsize=90mm\epsfbox{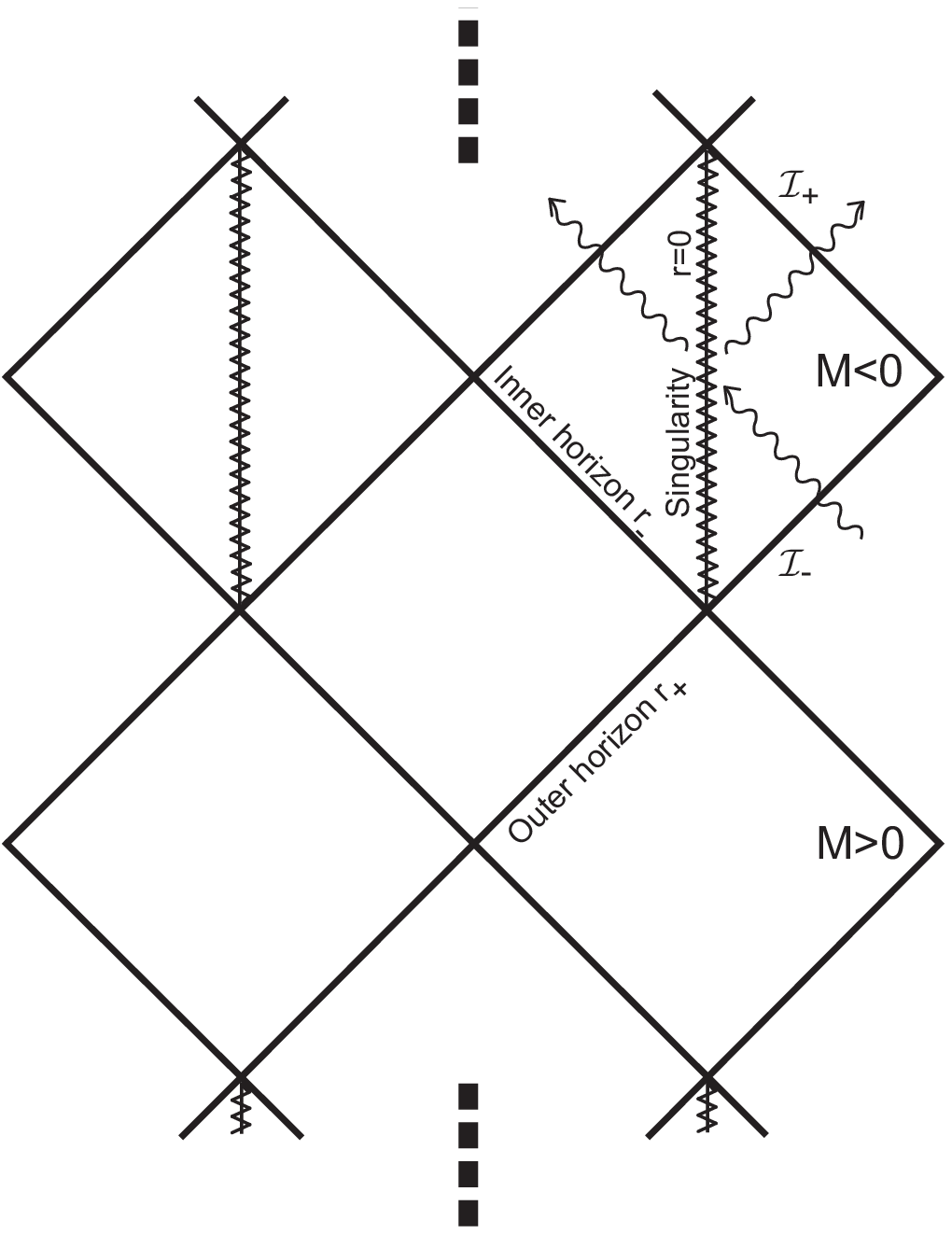}}
\medskip
\caption{The complete Penrose diagram for the (under-extremal
$|Q|<|M|$) Reissner-Nordstr\"{o}m background including the regions
with $r<0$ which are not drawn usually. These regions are glued to
``the other side'' of the time-like singularity, which is seen as
naked from that side. The wavy lines denote a typical scattering
experiment off the singularity. Waves from $\scri ^-$ are sent to
the singularity. Some of them are reflected back to $\scri ^+$
while the remainder is transmitted to $r>0$.} \label{PenroseRN}
\end{figure}

\begin{figure}
\centerline{\epsfxsize=120mm\epsfbox{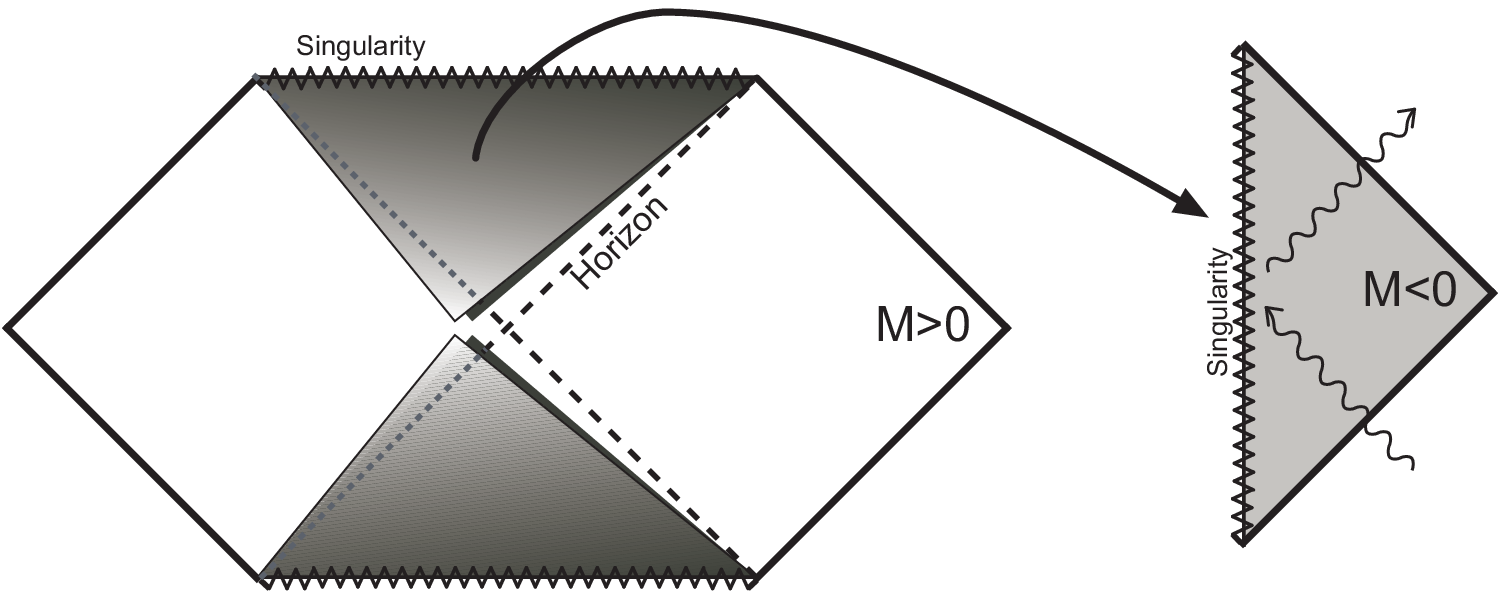}}
\medskip
\caption{The complete Penrose diagram for a Schwarzschild black
hole including the negative-mass regions which appear as a double
cover of the inside of the black (or white) hole. The
negative-mass regions are shown separately on the side, rotated by
90 degrees so that time flows in them upward as is conventional.
Here we will find no communication through the singularity.}
\label{PenroseSchw}
\end{figure}

Usually one considers only the $r>0$ part of the RN geometry to be
physical, but {\it we will find it natural to consider the $r<0$
region as well}. Note a crucial difference in this regard between
the $r=0$ singularities in Reissner-Nordstr\"{o}m and in
Schwarzschild: in Reissner-Nordstr\"{o}m $f$ has a second-order
pole at $r=0$,
\begin{equation}
f\cong {\frac{Q^{2}}{r^{2}},}  \label{fRN-lead}
\end{equation}
and the $t$ coordinate is time-like on both sides of $r=0$, while
in Schwarzschild $f$ has a first-order pole
\begin{equation}
f\cong -{\frac{2\,M}{r}}  \label{fSchw-lead}
\end{equation}
and the coordinates ``flip'' their space-like/time-like nature:
$t$ is space-like for small and positive $r$ and it is time-like
for negative $r$. For this reason (as well as for other reasons),
we will later find it natural to glue the positive-$r$ and
negative-$r$ patches of Reissner-Nordstr\"{o}m over the
singularity, but not in the Schwarzschild case. The Penrose
diagram of the maximally extended Reissner-Nordstr\"{o}m
spacetime, including the $r<0$ region, is given in figure
\ref{PenroseRN}. This is to be contrasted with figure
\ref{PenroseSchw} which provides the Penrose diagrams of the
regions $r>0$ and $r<0$ in the Schwarzschild case. These diagrams
-- and particularly the fact that the singularity is spacelike at
$r>0$ and timelike at $r<0$ -- indicate the difficulty of matching
the two regions at $r=0$.

\vspace{.5cm} \noindent {\it Kruskal-Szekeres coordinates}

Although the Schwarzschild coordinates are perfectly suitable for
describing the $r<0$ region and its matching to $r>0$, insight
into the ``completed Penrose diagram'' may be achieved by using
the Kruskal-Szekeres coordinates, as we now describe. (In a first
reading one may skip the text and only view the completed Penrose
diagrams).

The $(r,t)$ coordinates (or ``\Schw coordinates'') define several
metric patches, which are separated on the $r$ axis by $r=0,\,
r_-,\, r_+$ in \RN and $r=0,\, r_0$ in Schwarzschild. In order to
see how these patches, and especially the $r<0$ patch, are
connected one transforms to the Kruskal-Szekeres coordinates
\cite{Kruskal,Szekeres}, and later performs a conformal
transformation to obtain the Penrose diagram (in appendix
\ref{KSapp} we review the procedure). Here again we will find a
qualitative difference between the \RN singularity and
Schwarzschild.

In \KS coordinates $(U,V)$, one uses a function $g(r)$ to
implicitly define $r$ through \be
 -U\, V = g(r)~. \ee
This function is defined by \be
 g(r) := \exp \(\pm 2\, \kappa\, r^*\)~, \ee
in terms of the ``tortoise'' coordinate $r^*$: \be
 dr^* := {dr \over f}~. \ee
$\kappa$ is the surface gravity of the relevant horizon and the
choice of sign is correlated with the choice of sign for $U$ and
$V$.

\begin{figure}
\centerline{\epsfxsize=70mm\epsfbox{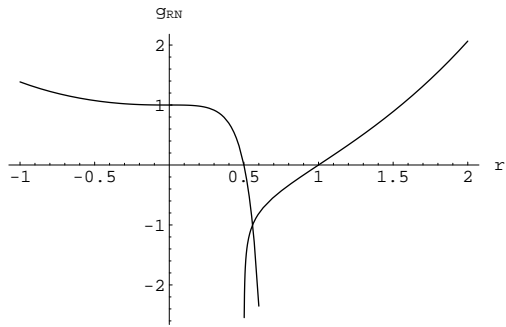}}
\medskip
\caption{The two functions $g_\pm(r)$ which determine $r$ in \KS
coordinates for \RN implicitly through $-U_\pm\, V_\pm = g_\pm
(r)$. $g_+$ is defined for the $r>r_-$, namely on the plus plane,
and $g_-$ is defined on $r<r_+$ namely on the minus plane. Note
that both functions are monotonic with range $(-\infty,+\infty)$
and hence there is always a unique solution for $r$. For $(-U_-\,
V_-)>1$ on the minus plane $r$ is negative. In drawing the figure
the values $r_-=.5, ~ r_+=1$ were used.}
 \label{gRNfigure}
\end{figure}

Let us look at the functions $g(r)$. For RN $g_\pm(r)$ are given
by (see figure \ref{gRNfigure}) \bea
 g_+(r) &=& \({r \over r_+}-1 \) \, \({r \over r_-}-1 \)^{-{\kappa_+ \over
\kappa_-}} \, \exp (2\, \kappa_+\, r) \non
 g_-(r) &=& \(1-{r \over r_-}\) \, \(1-{r \over r_+}\)^{-{\kappa_- \over
\kappa_+}} \, \exp
 (-2\, \kappa_-\, r) \eea
where $g_+$ covers the ``plus plane'' which includes the outer
horizon at $r_+: ~r_-<r<\infty$, and $g_-$ covers the ``minus
plane'' including the inner horizon at $r_-: ~-\infty<r<r_+$.
{}From the form of $g_-$ one realizes that in these coordinates
the $r<0$ region is naturally included and it lies at $(-U_-\,
V_-)
>1$, which is adjacent to the $0<r<r_-$ or $0 < -U_-\, V_- < 1$
region, and  as such we include it in the Penrose diagram (figure
\ref{PenroseRN}).

\begin{figure}
\centerline{\epsfxsize=70mm\epsfbox{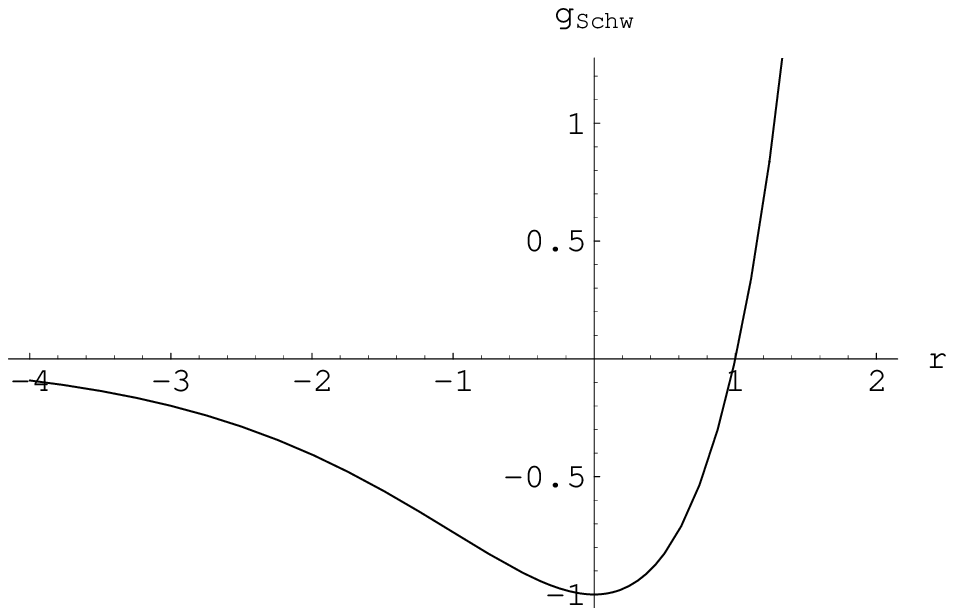}}
\medskip
\caption{The function $g_{\rm schw}(r)$ which determines $r$ in
\KS coordinates (for the \Schw black hole) implicitly through
$-U\, V = g_{\rm schw} (r)$. Note that while for $g_{\rm schw}>0$
there is a unique solution for $r$, for $-1<g_{\rm schw}<0$ there
are two solutions, one of them with negative $r$ which lead to a
double cover in \KS coordinates. Finally for $g_{\rm schw}<-1$
there are no solutions.}
 \label{gSchw-figure}
\end{figure}

For Schwarzschild on the other hand, we have \be
 g_{\rm schw}(r) = \({r \over r_0}-1 \)\, \exp(r/r_0). \ee
One notes that unlike the $g_\pm$ which were monotonic $g_{\rm
schw}(r)$ has a minimum at $g_{\rm schw}(r=0)=-1$ (see figure
\ref{gSchw-figure}), exactly because $f_{\rm schw}(r)$ changes
sign at $0$. Thus for $-U\, V<-1$ there is no solution for $r$
while in the region $-1<-U\, V<0$ there are two solutions one of
them with negative $r$. Hence, the $r<0$ region is naturally
included and becomes a second cover over the inside of the black
hole ($0<r<r_0$) which we incorporate into the Penrose diagram in
figure \ref{PenroseSchw}. This second cover differs from \RN where
the $r<0$ region was located side by side with the ``ordinary''
$r>0$ regions. This difference is in tune with our later
conclusion that for \RN one should glue the two regions over the
$r=0$ singularity while in \Schw they should be considered to be
disconnected.

\vspace{0.5cm}
 \noindent {\it Negative-energy content of the smeared singularity}

The electro-vacuum geometry constructed here has a curvature
singularity at $r=0$. One may consider modifying the metric
function in the very neighborhood of $r=0$, say at some $|r|\leq
r_{0}$, so as to make the geometry smooth. In this way one obtains
a wormhole smoothly connecting the positive-$r$ and the
negative-$r$ parts. However, any such smearing will imply the
presence of additional energy-momentum in the region $|r|\leq
r_{0} $, as would be determined by applying the Einstein operator
to the modified metric. The specific form of the additional
energy-momentum tensor will depend on the specific choice of
smearing. It is well known, however, that any wormhole of this
kind requires ``exotic'' matter fields, i.e. an energy-momentum
source which violates the weak energy condition. A simple way to
understand it is to consider a congruence of null rays propagating
inward radially. In this setting Raychaudhuri's equation tells us
that as long as the energy condition is satisfied ${d^2 \over
d\lambda^2} \log(A) \le 0$, where $A$ is the transverse area (of
the two-sphere) and $\lambda$ is an affine parameter along the
ray. Since $A$ (and $\log(A)$) is asymptotically decreasing in an
inward radial motion, but after going through the ``wormhole'' it
is again increasing, there must be an intermediate region where
the weak energy condition is violated.

We wish to evaluate the amount of negative energy required to
support this wormhole-shaped smeared geometry, particularly at the
limit $r_{0}\to 0$. To be more specific, we consider here a
thin-shell model \cite{Israel}. Namely, we directly match the
geometry at $r\geq r_{0}$ to that at $r\leq -r_{0}$, through a
thin shell located at $|r|=r_{0}$, for some $0<r_{0}\ll Q$. This
thin shell is a timelike hypersurface parameterized by the three
coordinates $(t,\theta ,\varphi )$. Note that the induced
three-geometry is continuous across this shell: $g_{\theta \theta
}$ and $g_{\varphi \varphi }$ only depend on $r^{2}$, and $g_{tt}$
(which originally has different values at the two sides) becomes
continuous after a trivial rescaling of the $t$ coordinate at e.g.
$r<-r_{0}$.

The shell's contribution to the energy-momentum tensor may be
expressed as \cite{Visser}
\begin{equation}
T^{\mu \nu }(x)=S^{\mu \nu }(x)\delta (\eta ) ~,
\end{equation}
where $S^{\mu \nu }$ is the shell's surface energy-momentum
distribution, and $\eta $ is the proper-length parameter along
radial lines of constant $t,\theta ,\varphi $, with $\eta =0$ at
the shell and $\eta >0$ at $r>r_{0}$. The surface distribution
$S^{\mu \nu }$ is determined from the jump in the extrinsic
curvature,
\begin{equation}
S_{j}^{i}=-\frac{1}{8\pi }\left( \left\langle K\right\rangle
_{j}^{i}-\delta _{j}^{i}\left\langle K\right\rangle
_{k}^{k}\right) .
\end{equation}
Hereafter $i,j,k$ run over the three coordinates $(t,\theta
,\varphi )$, and
\begin{equation}
\left\langle K\right\rangle _{j}^{i}\equiv
K_{j}^{i(+)}-K_{j}^{i(-)},
\end{equation}
where $K_{j}^{i(\pm )}$ is the shell's extrinsic curvature with
respect to the geometries at $r>r_{0}$ and $r<-r_{0}$,
respectively. (We are using here General-Relativistic units,
$G=c=1$.) The extrinsic curvature may conveniently be expressed in
terms of Israel's ``natural coordinates'' \cite{Israel}, which we
take here to be the three hypersurface coordinates $t,\theta
,\varphi$ and the above proper-length coordinate $\eta $:
\begin{equation}
K_{ij}^{(\pm )}=\frac{1}{2}\left[ \frac{\partial g_{ij}}{\partial
\eta } \right] _{(\pm )}=\frac{1}{2}\left[
g_{rr}^{-1/2}\frac{\partial g_{ij}}{
\partial r}\right] _{(\pm )}  ~, \label{extrinsic}
\end{equation}
with the derivatives (and $g_{rr}$) evaluated at the $(+)$ or
$(-)$ sides of the shell, respectively.

We are primarily interested here in the amount of (negative)
energy contributed by the shell, represented by $S_{t}^{t}$:
\begin{equation}
S_{t}^{t}=\frac{1}{8\pi }\left( \left\langle K\right\rangle
_{\theta }^{\theta }+\left\langle K\right\rangle _{\varphi
}^{\varphi }\right) =\frac{ 1}{4\pi }\left\langle K\right\rangle
_{\theta }^{\theta } ~.
\end{equation}
A straightforward application of Eq. (\ref{extrinsic}) yields
\begin{equation}
K_{\theta }^{\theta (\pm )}=\left[
\frac{\sqrt{1-2M/r+Q^{2}/r^{2}}}{r} \right] _{r=\pm r_{0}} ~,
\end{equation}
hence
\begin{equation}
\left\langle K\right\rangle _{\theta }^{\theta }=\frac{\sqrt{
1-2M/r_{0}+Q^{2}/r_{0}^{2}}+\sqrt{1+2M/r_{0}+Q^{2}/r_{0}^{2}}}{r_{0}}\,.
\end{equation}

Let $\rho $ denote the shell's energy density as measured by a
static observer:
\begin{equation}
\rho =-T_{t}^{t}=-\frac{1}{4\pi }\left\langle K\right\rangle
_{\theta }^{\theta }\delta (\eta )
\end{equation}
(recall that $T_{t}^{t}$ is invariant to a rescaling of $t$). The
total shell's energy is obtained by integrating over the 3-volume,
i.e. over $\eta $ and over the shell's area:
\begin{eqnarray*}
E_{shell} &=&4\pi r_{0}^{2}\int \rho d\eta =-r_{0}^{2}\left\langle
K\right\rangle _{\theta }^{\theta } \\
&=&-r_{0}\left( \sqrt{1-2M/r_{0}+Q^{2}/r_{0}^{2}}+\sqrt{
1+2M/r_{0}+Q^{2}/r_{0}^{2}}\right) .
\end{eqnarray*}

Finally, we calculate $E_{shell}$ at the limit where $r_{0}$
shrinks to zero:
\begin{equation}
E_{shell}^{0} \equiv \lim_{r_0 \to 0} E_{shell}=-2Q.
\end{equation}

\vspace{0.5cm}
 \noindent {\it Perspectives on the $r<0$ region}

The metric (\ref{RNmetric}) is invariant under the transformation
$r\to -r,~M\to -M,$ and hence we shall often refer to the $r<0$
region as the ``negative mass'' region. To an observer in the
asymptotically-flat region at $r<0$ the central object will appear
as one of negative mass, namely, a gravitationally-repelling
object.\footnote{This is also the situation in the Kerr geometry:
in the asymptotically-flat region $r<0$ the central object is
gravitationally repelling.}

We alert the reader that negative masses have very unusual
features:

\begin{itemize}
\item  A negative mass produces anti-gravity, namely it repels all
masses.

\item  As a result of Newton's second law $F=m\, a$, the
acceleration of a negative mass is reversed to the force acting on
it.

\item  As an immediate result of the two properties above, if one
were to place two bodies initially at rest, one with a negative
mass and the other with a positive mass, both will accelerate in
the same direction going from the negative mass to the positive
one (furthermore, if the two masses are of the same magnitude,
they will uniformly accelerate forever).

\end{itemize}

Due to these unusual features (and possibly others), negative
masses are often excluded (however, see
\cite{negative-supergravity}). We shall not attempt here to
provide a conclusive answer to this question. Instead, we will
take a pragmatic approach of exploration by allowing negative-mass
spaces in the current context and being alert to the appearance of
a problem such as a resulting inconsistency. We will not find any
such inconsistency in this paper, and so we regard this question
as remaining open.

In the (under extremal) RN black-hole case, $M>|Q|>0$, one can
reach the $r=0$ singularity either from the negative-mass
asymptotic region (i.e. from large negative $r$), or from $r \to
r_{-}$, the inner horizon. Looking at the Penrose diagram of the
maximally extended spacetime, figure \ref{PenroseRN}, one may
consider reaching the inner horizon by jumping into the black hole
from the ordinary (positive-mass) asymptotic region. The inner
horizon of the pure RN geometry is a perfectly-regular
hypersurface. However, it is known already for some time that the
inner horizon is unstable to perturbation and should become
singular \cite{Penrose,PI} in ``realistic'' charged black holes.
For this reason it was often believed that the inner horizon
cannot be crossed. Recent investigations, however, indicated that
the singularity at the inner horizon is weak, that is, the tidal
forces are too weak to harm physical objects
\cite{weak-singularity,burko,spinning}. Therefore, there is no
obvious reason to exclude objects falling into the black hole and
arriving at $r=0$ through the inner horizon. For conceptual
simplicity, however, we shall primarily consider here observers
arriving at $r=0$ from the negative-$r$ asymptotically-flat
region, namely, from beyond the singularity.

\begin{figure}
\centerline{\epsfxsize=55mm\epsfbox{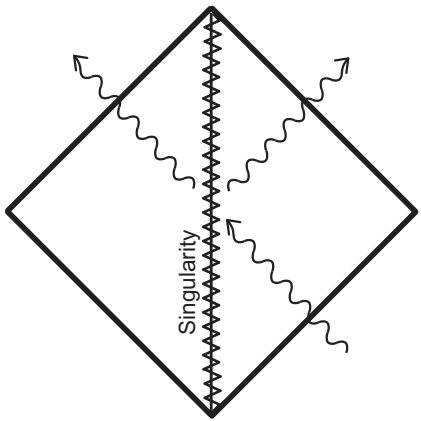}}
\medskip
\caption{The complete Penrose diagram for an over-extremal
($|M|<|Q|$) Reissner-Nordstr\"{o}m black hole. It is composed of a
positive-mass RN glued over the singularity to a negative-mass RN.
There are no horizons and the singularity is naked on both sides.}
\label{PenroseRN-ext}
\end{figure}

In the case $|Q|>M$ (figure \ref{PenroseRN-ext}) there are no
horizons, but simply two asymptotically-flat regions, one with
positive mass and the other with negative mass (in this case there
is no reason to transform into Kruskal-Szekeres -like
coordinates). In both regions there is a naked singularity at
$r=0$, and we glue the two patches there.

\vspace{0.5cm} \noindent {\it Geodesics}

We end this section with a review of particle motion in this
background, and a derivation of the effective potential for
particles (or geodesics) which we later compare with the effective
potential for waves. The geodesic equation for a particle of mass
$m$ can be derived from the Hamiltonian system
\begin{equation}
{\cal H}={\frac{1 }{2}}\, g^{\mu\, \nu}\, p_\mu\, p_\nu =
-{\frac{1 }{2}} \, m^2 ~~,
\end{equation}
where the Hamiltonian is automatically with respect to a multiple
of the proper time $\tau$. For a particle of charge $q$ one
replaces $p \to p-q\, A$.

Since ${\cal H}$ is independent of $t$ and $\,\phi $ we have the
constants of the motion $E:=-p_{t},\,l:=p_{\phi }$. The motion is
then described by
\begin{eqnarray}
0 &=&\dot{r}^{2}+V_{{\rm eff}}^{{\rm geodesic}}(r)~,  \nonumber \\
V_{{\rm eff}}^{{\rm geodesic}}(r) &=&\left({\frac{l^{2}}{r^{2}}}\,
+m^{2}\right)\,f(r)-\left(E-{\frac{q\,Q}{r}}\right)^{2}~.
\label{Veff-geod}
\end{eqnarray}
In particular, around $r\simeq 0$, $V_{{\rm eff}}\sim
{\frac{l^{2}Q^{2}}{r^{4}}}$ for $l>0$ [and $V_{{\rm eff}}\sim
(m^{2}-q^{2}){\frac{Q^{2}}{r^{2}}}$ for $l=0$]. Note that the
effective potential given here does not have the ordinary units of
potential, but it has the advantage of being general and valid
both for massive and massless particles. In order to get the
physical potential in a Newtonian approximation one needs to
rescale $V_{{\rm eff}}$ by an appropriate power of $m$.

\section{Waves are smooth}
\label{smooth-section}

\subsection{Set-up}

A spacetime is usually defined to be singular when it is not
geodesically complete, namely with respect to point-particle
probes. Yet, all the fundamental interactions are formulated for
field theories, where particles are derived concepts constructed
from appropriate wave packets. Moreover, the mashing of Quantum
Mechanics and Special Relativity requires the quantum theory of
fields. Thus it is natural to probe spacetimes with waves rather
than point-particles.

Here we will confine ourselves, as a first step, to scalar fields.
Initially we will further assume a minimally-coupled free field
with no mass or charge, and later these assumptions will be
relaxed and shown not to affect the results.

When should we consider a spacetime to be wave-regular? A central
objective of physics is to provide predictions for the results of
experiments, or a well-defined time evolution, and it is exactly
this predictability which is at risk in the presence of a
singularity. We would therefore consider a spacetime to be
wave-regular if the time evolution is unique for any physically
acceptable initial conditions (to be discussed below). Technically
the issue is whether one can supply boundary conditions at the
singularity such that the time evolution is regular and whether
these are unique.

In ``normal'' (i.e. non-singular) situations there are two
different set-ups for initial conditions: (i) the Cauchy
formulation, in which the field and its time derivative are
specified on a spacelike hypersurface; and (ii) the characteristic
formulation, in which the field is specified on two null
hypersurfaces. This second formulation has a special, widely used
variant (iia) in which the null hypersurfaces are taken to be at
the null past boundaries of the region of spacetime under
consideration -- e.g. past null infinity (for an
asymptotically-flat region), and a past horizon (for an
asymptotically-flat region outside a black hole).

In our problem -- formulating the time evolution of the field at
the two sides of the $r=0$ singularity -- the Cauchy formulation
(i) has a potential problem, because any spacelike initial
hypersurface must cross the $r=0$ singularity. It is a priori
unclear what are exactly the ``physically-acceptable initial
conditions'' near $r=0$. On the other hand, the characteristic
formulation (iia) is free of this potential problem. The two null
hypersurfaces which naturally suggest themselves for this
formulation are past null infinity ($\scri^-$) of the negative-$r$
universe, and the inner horizon $r_{-}$. These two hypersurfaces
nowhere intersect the singularity, or even get close to it.

We shall therefore base our analysis of time evolution on this
characteristic formulation. That is, we probe the spacetime by
wave-packets prepared and sent in from the asymptotic regions far
away from the singularity. Indeed, it is physically reasonable to
characterize the singularity by its response to any possible
scattering experiment. Motivated by these considerations, we shall
define a spacetime (or a region of spacetime) to be {\it
wave-regular} if the time evolution is unique for generic initial
conditions specified at the relevant past null boundaries of this
(region of) spacetime.

While the characteristic formulation (ii) avoids the issue of
putting constraints on initial conditions near the singularity,
which is indeed non-trivial in hindsight, it does require a global
point of view, whereas our resolution is essentially local. This
local property is better understood in the Cauchy formulation (i)
and hence we will briefly review both formulations in this
section. Actually one could define also a hybrid of these methods
which keeps both locality and some isolation from the singularity:
namely, a formulation (ia) in which the field and its derivatives
are supplied on a spacelike hypersurface but are required to
vanish in a neighborhood of the singularity.~\footnote{This
variant is slightly more restrictive than just a re-formulation of
the initial-value setup, as it demands that the field strictly
vanishes at some open neighborhood. Nevertheless it is still
sufficiently general to allow for physically-meaningful scattering
experiments.}

In what follows we shall show that both the $-\infty < r < r_-$
piece of the RN spacetime and negative-mass \Schw are indeed
wave-regular. We shall define first the time evolution in a
straightforward manner in the Cauchy formulation, and then after
defining a canonical form for the radial equation and the
effective potential we shall state the characteristic formulation.
A treatment of the interesting subtleties in the Cauchy
formulation will be deferred to section \ref{conditions-section}.

\subsection{Separation of variables}

Let us proceed by writing down the wave equation for a minimally
coupled massless scalar field $\phi $ in the RN (or Schwarzschild)
background (\ref {RNmetric}). After separating the angular
variables \be
 \phi(r,t,\Omega) = \phi_l(r,t)\, Y_{lm}(\Omega)~, \ee
 where $\Omega$ denotes the angular variables~\footnote{The notation
$\Omega$ for the angular variables is convenient for the
generalization to higher dimensions where there are more angular
variables and there is no need to explicitly define them.}
 $(\theta,\, \varphi)$, one has
\begin{equation}
0=\Box \phi =\left[ -f^{-1}\,{\partial
}_{tt}+{\frac{1}{r^{2}}}\,{\partial }_{r}\,f\,r^{2}\,{\partial
}_{r}-{\frac{l\,(l+1)}{r^{2}}}\right] \,\phi _{l}~.
\label{waveeq1}
\end{equation}
The wave equation (\ref{waveeq1}) is singular at $r=0$: after
multiplying the equation by $f$ to extract ${\partial }_{tt}$, the
part containing e.g. the second-order $r$ derivative becomes
$f^{2}\,{\partial }_{rr}$, and $f^{2}$ diverges as $r^{-4}$. This
is not surprising since we know that $r=0$ is a curvature
singularity.

Still, we may attempt a straight-forward separation of the time
variable,
\begin{equation}
\phi _{l}(r,t)=\phi _{\omega l}(r)\,{\rm exp}(i\,\omega \,t)~,
\label{basicseparation}
\end{equation}
getting \bea
 0 &=& \Box_{\omega l}\, \phi_{\omega l} \non
 \Box_{\omega l} &:=& f^{-1}\, \omega^2+
 {\frac{1}{r^{2}}}\,{\partial }_{r}\,f\,r^{2}\,{\partial }_{r}-
 {\frac{l\,(l+1)}{r^{2}}}~. \eea
It will be convenient to normalize $\Box_{\omega l}$ in two
different ways: if we want to have a unit coefficient for
${\partial }_{rr}$ then we work with $f^{-1}\, \Box_{\omega l}$
while $f\, \Box_{\omega l}$ achieves a unit coefficient for
$\omega^2$. Indeed, writing down $f^{-1}\, \Box_{\omega l}$
explicitly, we have \bea
 0 &=& f^{-1}\, \Box_{\omega l}\, \phi_{\omega l} =\non
 &=& \left[ {\frac{1}{f\,r^{2}}}\, \partial_{r}\,f\, r^2\, \partial
_r-{\frac{l\,(l+1)}{f\,r^2}} + f^{-2}\, \omega ^{2}\right]
\,\phi_{\omega l} ~~, \label{separation-of-var} \eea and at this
point we need to discuss the Reissner-Nordstr\"{o}m case and the
Schwarzschild case separately.

\vspace{.5cm} \noindent {\it Reissner-Nordstr\"{o}m }

Since $f_{RN}\,r^{2}\cong Q^{2}$ is regular at $r=0$, the
variable-separated equation (\ref{separation-of-var}) is
surprisingly completely regular. This observation is at the heart
of our proposed resolution. To leading order it is given by
\begin{equation}
0=\left[ (1+\dots )\,{\partial }_{rr}+O(r^{0})\,{\partial
}_{r}-(1+\dots )\,{\frac{l\,(l+1)}{Q^{2}}}+(1+\dots
)\,{\frac{r^{4}}{Q^{4}}}\,\omega ^{2}\right] \phi _{\omega}~,
\label{rnreg}
\end{equation}
where ellipses always denote corrections which are higher order in
$r$, and from now we will often write $\phiom$ instead of
$\phi_{\omega l}$ suppressing the index $l$ for clarity. Note that
the equation is not only non-singular but also analytic (namely,
$\phi _{\omega ,rr}$ is an analytic function of $\phi _{\omega
},\phi _{\omega ,r}$ and $r$) at $r=0$, for any $l$ and $\omega $.

Since all the solutions for this equation are smooth at $r=0$, a
natural way to define the time evolution suggests itself: given an
initial wave packet (which propagates from either $\scri^-$ and/or
 the inner horizon) decompose it into radial eigen-functions $\phi
_{\omega }(r)\,$. Since the eigen-functions $\phi _{\omega }(r)$
are smooth at $r=0$, the time evolution defined above will be
smooth there. Consequently, parts of the wave packet will cross
this point. It is therefore essential to include both sides of the
$r=0$ singularity in our analysis.

In equations: given initial conditions $\phi_i(r):=\phi(r,t=0)$
and $\dot{\phi}_i(r):=\dot{\phi}(r,t=0)$ (we chose here the
initial hypersurface $t=0$ without loss of generality), one
determines the decomposition $\phi_+(k),\, \phi_-(k)$ that
satisfies \bea
 \phi_i(r) &=& \int \( \phi_+(k) + \phi_-(k)\, \)\, \phi_k(r)\, dk \non
 \dot{\phi}_i(r) &=& \int i \omega\, \( \phi_+(k) - \phi_-(k)\, \)\,
\phi_k(r)\, dk
 \eea
where here $\omega:=|k|$, and $\phi_{k}$ denotes two independent solutions
(one for positive $k$ and one for negative $k$)
of the wave equation $\Box_{\omega l}=0$ (with $\omega=|k|$).
 \footnote{$\phi_{k}$
are usually chosen such that $\phi_k \simeq \exp (i\, k\,
r^*\,)/r$ at large $r^*$ (large negative $r$).}
 Now the time evolution is defined naturally to be \footnote{This amounts to
 taking the radial function $\phi _{\omega l}(r)$
 of Eq. (\ref{basicseparation}) to be
 $\phi_+(k=+\omega)+\phi_+(k=-\omega)$ for $\omega>0$ and
 $\phi_-(k=+\omega)+\phi_-(k=-\omega)$ for $\omega<0$.}
 \be
 \phi(r,t) = \int \( \phi_+(k)\,  \exp(+i\, \omega\, t)
 + \phi_-(k)\,  \exp(-i\, \omega\, t)\, \) \phi_k(r)\, dk ~~.\ee

In the characteristic formulation the decomposition into modes may
be done by Fourier-decomposing the waves coming from $\scri^-$ or
from the inner horizon in ${\rm exp}(i\,\omega \,v)$ or ${\rm
exp}(i\,\omega \,u)$, respectively, where $u$ and $v$ are the two
Eddington-like null coordinates, and associating a certain radial
function $\phi _{\omega }(r)$ with each Fourier component. This
will be described in more detail in subsection
\ref{char-subsection}.

At first it may look as if the singular partial differential
equation (\ref {waveeq1}) has turned completely regular. Later we
will see, however, that some imprint of the singularity remains.
This is expressed both in the unusual large-$\omega$ limit of the
transmission amplitude through $r=0$ (see section
\ref{cross-section-section}) and in the constraints for Cauchy
initial conditions at $r \sim 0$ (see section
\ref{conditions-section}).

\vspace{.5cm} \noindent {\it Schwarzschild }

Using the leading behavior of $f_{{\rm Schw}}$ at $r=0$
 (\ref{fSchw-lead}), the variable-separated equation
(\ref{separation-of-var}) is given to leading order by
\begin{equation}
0=\left[ (1+\dots )\,{\partial }_{rr}+{\frac{(1+\dots
)}{r}}\,{\partial } _{r}+(1+\dots
)\,{\frac{l\,(l+1)\,}{2\,M\,r}}+(1+\dots
)\,{\frac{r^{2}}{4\,M^{2}}}\,\omega ^{2}\right] \phi _{\omega } ~.
\label{Schw-var-sep}
\end{equation}
One notices that this equation is not regular anymore, but rather
the $\partial_r$ and the $l(l+1)$ terms are singular. Actually it
falls into the so called ``regular-singular'' category of mild
singularities (see appendix \ref{reg-sing-app} for a rudimentary
review).

Looking at Eq. (\ref{Schw-var-sep}) one realizes that the
characteristic exponents (defined by the attempted leading-order
solution $\phi _\omega \cong r^{\rho }$) are $\rho
_{1}=\rho_{2}=0$, and thus the two independent solutions behave as
$\phi _{1 \omega}\sim r^{0},~\phi _{2 \omega}\sim \log (r)$ at
$r\sim 0$. So unlike Reissner-Nordstr\"{o}m one of the solutions
is singular, and moreover does not have a single-valued
continuation to negative $r$, and there is no sense in gluing
together positive $r$ to negative $r$. It is perhaps fortunate
that this is the case, since gluing the two sides would require
dealing with the ``flip'' in the nature of $(r,t)$ from space-like
to time-like (and an arbitrary choice of the direction of the
arrow of time). Still we find it remarkable that one of the
solutions does turn out to be regular. This half-regularity allows
a natural definition of the time evolution, as we shortly
describe, where this time there is no transmission through $r=0$.

The situation here is analogous to the relation between a free
field $\phi $ on the infinite line $-\infty <r<+\infty $ to a free
field on the semi-infinite line $0<r<+\infty $, which could be
e.g. the radial coordinate in some higher dimension, with
Dirichlet or Neumann boundary condition at $r=0$. For the infinite
line there are two $\phi_{\omega }$ solutions for each $\omega$,
which allow a decomposition of wave packets arriving
simultaneously from both $-\infty$ and $+\infty$. On the other
hand, for the semi-infinite line the boundary condition selects a
single $\phi _{\omega }$ solution for each $\omega $. This still
allows for a decomposition of a wave packet, which this time can
arrive only from $+ \infty $. The situation for
Reissner-Nordstr\"{o}m is similar to the infinite line, while
negative-mass Schwarzschild is similar to the semi-infinite line
(we consider the $r<0$ side of the Schwarzschild singularity since
there $t$ is time-like and one can consider scattering
experiments).

Thus we naturally define the time evolution for negative-mass
Schwarzschild by decomposing the initial conditions (defined only
at $r=-\infty $) into the regular eigen-functions $\phi _{1
\omega}$ alone and then proceeding as usual to define the time
evolution. Since $\phi _{1 \omega}$ is smooth at $r=0$, the time
evolution will be smooth there too. As it turns out it was found
already in \cite{IshibashiHosoya} by somewhat formal arguments
that there exists a unique time evolution for negative-mass
Schwarzschild, which is probably the same as the one we just
explicitly described. For RN on the other hand we differ as we get
wave-regularity after considering both sides of the singularity,
while they do not.

\sbsection{Perspectives on the boundary conditions for \Schw}

In the \Schw case we simply impose the boundary condition that the
scalar field be regular at $r=0$.
This does not conform with either the Dirichlet or Neumann b.c.;
namely, neither $\phi$ nor its radial derivative vanishes at $r=0$.
In fact, the Dirichlet or Neumann b.c. are both inconsistent with
the asymptotic behavior near the singularity (due to the presence of a
divergent mode).

Yet, there is a physically appealing procedure
to obtain our regularity b.c.
from the Dirichlet or Neumann b.c. (or a mixture).
We can cutoff space at some $r_1<0$ close to the
singularity, and then take the limit $r_1 \to 0$. Since all modes are
well-behaved at $r=r_1$,
we are free to impose either Dirichlet or Neumann b.c. there.
Let us denote as before the
solution regular at $r=0$ by $\phi_{1 \omega}$, and some other independent and
necessarily singular solution by $\phi_{2 \omega}$. We are looking
for a linear combination of the two, $\phiom = a\, \phi_{1 \omega}
+ b\, \phi_{2 \omega}$, that will satisfy the boundary conditions at
$r_1$. If we impose Dirichlet we need \be
 0=\phiom(r_1) = a\, \phi_{1 \omega}(r_1) +b\,
 \phi_{2 \omega}(r_1) ~. \ee
 As we take $r_1 \to 0$ $\phi_{1 \omega}$
 remains finite while $\phi_{2 \omega}$
diverges with a log singularity, and hence we must take $b=0$,
namely the ``regular solution b.c.''

If we choose Neumann conditions instead, we note that again while
$\phi_{1 \omega}'$ remains finite $\phi_{2 \omega}'$ diverges with
a $1/r$ singularity and hence we are again forced to take $b=0$.
Altogether we find that in both cases as $r_1 \to 0$ we are left
with a unique b.c., namely our ``regular b.c.''

\subsection{Normal form of the radial equation}

\label{normal-form-subsection}

Transforming to a new radial coordinate (the so called ``tortoise
coordinate'') $r^*(r)$ defined by
\begin{equation}
dr^{*}={\frac{dr}{f(r)},} \label{def-rstar}
\end{equation}
and to a new field variable
\begin{equation}
\phi ^{*}:= r\phi  \label{defn-g}
\end{equation}
(and correspondingly $\phi _{\omega }^{*}:= r\phi _{\omega }$),
the wave equation (\ref{waveeq1}) and the radial equation
(\ref{separation-of-var}) take the standard forms
\begin{equation}
f^{-1}\, \left[ {\partial }_{tt}-{\partial }_{r*r*}+V_{{\rm
eff}}(r^*)\right] \phi^*=0 \label{normal-full}
\end{equation}
and
\begin{equation}
\left[ -{\partial }_{r*r*}+V_{{\rm eff}}(r^*)-\omega ^{2}\right]
\phi _{\omega }^{*}=0.  \label{normal-radial}
\end{equation}
Here the effective potential $V_{{\rm eff}}$ is given by $V_{{\rm
eff}}^{\rm geodesic}$, the effective potential for particles,
with the substitutions $l^2 \to l(l+1)$ and $E \to 0$,
\footnote{in principle $E$ is to be now replaced by $\omega$, but
the latter already appears explicitly in Eq.
(\ref{normal-radial}). Recall also that we assume at this stage a
neutral massless scalar field, hence $m=q=0$.} plus an extra term
$\Delta V_{{\rm eff}}$:
\begin{equation}
V_{{\rm eff}}(r^*)=\Delta V_{{\rm eff}}+f\,{\frac{l(l+1)}{r^{2}},}
\label{Veff}
\end{equation}
where
\begin{equation}
\Delta V_{{\rm eff}}(r^*)={\frac{f}{r}}\,{\partial }_{r}\,f ~.
\label{DeltaVeff}
\end{equation}

We find it instructive to explore the leading terms in the radial
equation at $r=0$, when this equation is expressed in terms of
$r^*$ rather than $r$.

We start by looking at the Reissner-Nordstr\"{o}m case. To leading
order near $r\sim 0$
\begin{equation}
r^*\cong {\frac{r^{3}}{3\,Q^{2}},}  \label{r*-r-at-0}
\end{equation}
and hence \be V_{{\rm eff}}\cong -2{\frac{Q^{4}}{r^{6}}}\cong
-{\frac{2}{9\,r^{*~2}}~.} \label{two-ninths}\ee
 Thus even though the wave equation in the form (\ref{separation-of-var})
is smooth, the effective potential in the normal form is unbounded
from below; hence we will study later (in section
\ref{cross-section-section}) its response to high-frequency waves,
where all features of the potential other than its leading term
$\propto r^{*~-2}$ are irrelevant.
 Note that such potentials of the type $V=-c/r^{*~2}$ are special (see for instance \cite{LL35}).
Considering a larger
 class of potentials $V=-c/r^{*~\alpha}$ with $c,\alpha>0$, then $\alpha=2$ is a
critical value describing a conformal quantum mechanics
 \footnote{if we replace $\omega^2 \to \omega$ in (3.14).},
 namely $c$ is dimensionless. For $0<\alpha<2$ the spectrum is bounded from
below even though the  classical potential is not, while for
$\alpha>2$ the spectrum is not bounded from below. For the
critical value $\alpha=2$ there is a critical value for $c$,
$c=1/4$. For $c<1/4$ the potential is still mild enough to have a
spectrum bounded from below (actually at zero),
 while for $c>1/4$ it is not. So the numerical coefficient $c=2/9$ in (\ref{two-ninths}) is
important. We will see later (in section \ref{generalizations})
that $c$ varies over our examples (it depends on the number of
dimensions, but always $c<1/4$ in the charged case), and that it
determines the high-energy transmission properties of the
singularity.

At $r^{*}=0$ the radial equation (\ref{normal-radial}) has a
regular singularity, and we may look for the characteristic
exponents $\rho $ defined by $~\phi _{\omega }^{*}\cong r^{*~\rho
}$. These satisfy the characteristic equation
\[
\rho \,(\rho -1)+{\frac{2}{9}=0,}
\]
leading to the two roots
\begin{equation}
\rho _{1,2}={\frac{1}{3}},\,{\frac{2}{3},}
\end{equation}
which are independent of $\omega $. Translating back from $r^{*}$
to $r$ using the transformation (\ref{defn-g},\ref{r*-r-at-0}) we
get the exponents $r^{0}\,$and $r^{1},$ exactly as expected for a
smooth second-order equation.

For Schwarzschild one may repeat the calculations based on the
leading behavior of $f_{{\rm Schw}}$ (\ref{fSchw-lead}) and find
\be
 r^{*}\cong -{\frac{r^{2}}{4M},} \ee leading to \be
 V_{{\rm eff}}\cong -4{\frac{M^{2}}{r^{4}}}\cong -{\frac{1}{4\,r^{*~2}}.}
\ee
 Thus we have here the critical value $c=1/4$ (independent of the
spacetime dimension, see section \ref{generalizations}).

Attempting $\phi _{\omega }^{*}\cong r^{*~\rho }$ one obtains the
characteristic equation
\[
\rho \,(\rho -1)+{\frac{1}{4}=0,}
\]
which yields \be \rho _{1}=\rho _{2}={\frac{1}{2}.} \ee The
equality of the two exponents implies the presence of a log term
in one of these solutions. Again translating back the
characteristic exponents to the $r$ coordinate we get the
$r^{0},\,\log (r)$ behavior which was derived above from Eq. (\ref
{Schw-var-sep}).

\subsection{Characteristic formulation}
\label{char-subsection}

Having defined the ``tortoise'' coordinate $r^*$ (\ref{def-rstar})
and the effective potential (\ref{Veff}) we can now return to
spell out in detail the definition of time evolution in the
characteristic formulation.

The coordinate $r^*$ approaches $-\infty $ as $r\to -\infty $ and
$+\infty $ as $r\to r_{-}$. Notice that the effective potential
decays as $r^{-2}\simeq $ $r^{*~-2}$ at large negative $r$, and as
$f\propto r-r_{-}$ (which is exponentially small in $r^*$) at the
inner horizon.
Therefore, in each of these asymptotic boundaries the radial
function is a superposition of the two asymptotic solutions
\begin{equation}
\phi^{*}_\omega \cong {\rm exp}(+ i\, \omega \,r^*) \, \mbox{  and  } \,
\phi^{*}_\omega \cong {\rm exp}(- i\, \omega \,r^*)~.
\label{gen-asymptotic}
\end{equation}
Defining the Eddington-like coordinates
\[
v:= t-r^*\,,\qquad u:= t+r^*,
\]
the above two asymptotic solutions at both edges $r^*\to \pm \infty $
read, for a particular $\omega $ component of $\phi^*$,
\[
\phi^* \cong {\rm exp}(i\,\omega u)\,\, \mbox{  and  } \,\,\phi^*
\cong {\rm exp}(i\,\omega v)~,
\]
respectively. The non-decomposed field at both asymptotic edges
takes the form
\[
\phi^*(r^*\to \pm \infty )\cong f_{\pm }(v)+g_{\pm }(u),
\]
where $f_{\pm }(v)$ and $g_{\pm }(u)$ are arbitrary functions (the
``$\pm $'' stands for the two asymptotic regions $r^*\to \pm
\infty $, i.e. to the inner horizon and to large negative $r$). In
the characteristic initial-value formulation one needs to specify
the ingoing component $f_{-}(v)$ at $\scri^-$ and the component
$g_{+}(u)$ coming from the inner horizon. These two functions
should uniquely determine the time evolution in the relevant piece
of spacetime. This evolution is defined as follows. First,
Fourier-decompose these two initial functions:
\begin{eqnarray*}
f_{-}(v) &=&\int f_{v}(\omega _{v})\,{\rm exp}(i\,\omega
\,_{v}v)d\omega
_{v}, \\
g_{+}(u) &=&\int g_{u}(\omega _{u})\,{\rm exp}(i\,\omega
\,_{u}u)d\omega _{u}.
\end{eqnarray*}
Next, we define a set of basis solutions $\phi _{(\omega
_{v},0)}^{*}(r^{*},t)$ and $\phi _{(0,\omega _{u})}^{*}(r^{*},t)$
(for each $\omega _{v}$ and $\omega_{u}$) as follows: $\phi
_{(\omega _{v},0)}^{*}$ is the solution evolving from the
characteristic initial conditions
\[
f_{-}(v)={\rm exp}(i\,\omega \,_{v}v)\,,\qquad g_{+}(u)=0.
\]
Similarly, $\phi _{(0,\omega _{u})}^{*}$ is the solution evolving
from the characteristic initial conditions
\[
f_{-}(v)={\rm 0}\,,\qquad g_{+}(u)={\rm exp}(i\,\omega \,_{u}u).
\]
The two functions $\phi _{(\omega _{v},0)}^{*}$ and $\phi
_{(0,\omega _{u})}^{*}$ are constructed from the radial functions
$\phi _{\omega }^{*}(r^*)$ (with $\omega=\omega_v$ and
$\omega=\omega_u$, respectively); see appendix \ref{char-appendix}
for details.

Once $\phi _{(\omega _{v},0)}^{*}$ and $\phi _{(0,\omega
_{u})}^{*}$ are defined, the time evolution of the field $\phi^*$
emerging from the characteristic initial data $f_{-}(v)$ and
$g_{+}(u)$ is simply given by
\begin{eqnarray*}
\phi^*(r^*,t) &=&\int f_{v}(\omega _{v})\,\phi _{(\omega
_{v},0)}^{*}(r^*,t)d\omega _{v} \\
&&+\int g_{u}(\omega _{u})\,\phi _{(0,\omega
_{u})}^{*}(r^*,t)d\omega _{u}.
\end{eqnarray*}
We conclude that the unique extension of the radial functions
$\phi _{\omega }^{*}(r^*)$ across the $r=0$ singularity naturally
leads to a unique time
evolution for any set of characteristic initial functions $f_{-}(v)$ and
$g_{+}(u)$.

\section{Transmission cross section in \RN}

\label{cross-section-section}

Having demonstrated a resolution of the $r=0$ singularity we turn
in this section to examine its nature through scattering. Here we
shall only consider the charged (Reissner-Nordstr\"{o}m) case.
Full reflection from the negative-mass Schwarzschild singularity
will be considered together with the $d>4$ cases at the end of
subsection \ref{highd}. We focus here on the transmission of
high-frequency incoming waves through the $r=0$ singularity (from
now on we will use ``high-energy'' to mean ``high-frequency'' by a
slight abuse of language originating from the quantum theory).

 We first study the effective potential $V_{{\rm
eff}}$, and especially its asymptotic form at small $r$. Then we
compute the transmission and reflection coefficients for fixed $l$
in the high-energy limit. Finally we combine that result with the
high-$l$ limit, where the partial cross section vanishes. We sum
over $l$ and obtain the $\omega $ dependence of the (high-energy)
total cross section for transmission.

\subsection{General features of the effective potential}

Consider first the qualitative behavior of the effective potential
$V_{{\rm eff}}$. For concreteness we shall consider here a wave
propagating from past null infinity of the negative-$r$ universe
towards $r=0$. At large $|r|$, we have the usual centrifugal
barrier
\begin{equation}
V_{{\rm eff}}\cong \frac{l(l+1)}{r^{2}}\qquad \qquad (|r|\gg M).
\label{Vlarger}
\end{equation}
At small $|r|$, the potential takes the asymptotic form
\begin{equation}
V_{{\rm eff}}\cong
l(l+1)\frac{Q^{2}}{r^{4}}-2\frac{Q^{4}}{r^{6}}\qquad \qquad
(|r|\ll Q).  \label{Vsmallr}
\end{equation}
Here we only keep the leading order in $r/Q$. We do keep, however,
the term $ \propto r^{-4}$ in the right-hand side because it is
proportional to $l(l+1)$. This term will become important for
partial waves of sufficiently large $l$, which are involved in the
calculation of the total high-energy transmission cross-section
(see below).

The contribution of any particular $l$ to the total high-energy
cross-section is suppressed by the kinematic factor $k^{-2}$ in
Eq. (\ref {total-cross-section}) below (while it will turn out
that $\sigma \gg k^{-2}$). Therefore, at high energy the total
cross section is dominated by partial waves with $l\gg 1$, which
we shall assume hereafter. Such waves of relatively small or
moderate energy will be reflected already at the large-$|r|$
centrifugal barrier (\ref{Vlarger}). Only waves with sufficiently
high energy, $\omega >l/Q,$ will penetrate into the central region
$|r|\ll Q$. The scattering features of these high-energy waves
will be dominated by the small-$|r|$ potential (\ref{Vsmallr}).
This potential has a peak value (hereafter we often replace
$l(l+1)$ by $l^{2}$, which is justified because $l\gg 1$)
$$
V_{\max }\cong \frac{l^{6}}{27Q^{2}},
$$
located at
$$
|r|\cong r_{peak}:= \sqrt{3}Q/l\,.
$$
It acts as a repulsive barrier ($\propto r^{-4}$) at $Q\gg
|r|>r_{peak}$, and as a potential well ($\propto -r^{-6}$) at
$|r|<r_{peak}$.

The above discussion immediately suggests that partial waves with
$\omega \ll \omega _{peak}$, where
\begin{equation}
\omega _{peak}:= \sqrt{V_{\max }}\cong \frac{l^{3}}{\sqrt{27}Q}\,,
\label{wpeak}
\end{equation}
will be fully reflected by the potential barrier at
$|r|>r_{peak}$. However, partial waves with $\omega \gg \omega
_{peak}$ will predominantly feel the potential well $\propto
-r^{-6}$ at $|r|<r_{peak}$. These waves will be partially
transmitted, as we analyze in the next subsection.

For later convenience we transform the small-$|r|$ potential from
$r$ to $r^*$, using $r^{3}\cong 3\,Q^{2}\, r^*$:
\begin{equation}
V_{{\rm eff}}\cong
\frac{l^{2}}{Q^{2/3}(3r^*)^{4/3}}-{\frac{2}{9\,r^{*~2}}} \qquad
\qquad (r\ll Q).  \label{Vsmallr*}
\end{equation}
The peak is located at
$$
|r^*|\cong r^*_{peak}:= \sqrt{3}Q/l^{3}\,.
$$
Although we are primarily considering here a wave coming from the
negative-$r$ past null infinity, the case of a wave propagating
from $r_{-}$ towards $r=0$ may be treated in exactly the same
manner. Note that the small-$|r|$ potential (\ref{Vsmallr}) or
(\ref{Vsmallr*}) is even in $r$ or $r^*$, respectively.
Consequently the high-energy transmission amplitude for partial
waves is the same for both cases. (Obviously the transmission
probability is exactly the same, even for finite $\omega $.)

\subsection{Amplitudes at high energy and fixed $l$}

\label{trasnmission-reflection}It is instructive to look at the
scattering in the high-energy limit. Normally, the potential is
bounded, hence the transmission amplitude approaches $1$ and the
reflection vanishes at this limit. However, here we have an
unbounded (attractive) potential near $r=r^{*}=0$, which leads to
partial reflection even at the high-energy limit.

For fixed $l$ and high energy, $\omega \gg \omega _{peak}$, the
potential is dominated by
$$
V_{{\rm eff}}\cong -{\frac{2}{9\,r^{*~2}}\equiv }V_{{\rm
eff,sing}}\,.
$$
We therefore need to solve the scattering problem
\begin{equation}
0=[-{\partial }_{r*r*}+V_{{\rm eff,sing}}(r^*)-\omega ^{2}]\phi
^{*}. \label{V-singular}
\end{equation}
Passing to the dimensionless variable
 \footnote{The negative sign
in the definition of $x$ was introduced in order for the incoming
wave to propagate from large positive $x$ values (null infinity)
towards smaller $x$ values.}
\begin{equation}
x=-r^{*}\,\omega ,  \label{xr*}
\end{equation}
the equation is transformed into
\begin{equation}
0=[-{\partial }_{xx}-{\frac{2}{9\,x^{2}}}-1]\,\phi ^{*}.
\end{equation}
Its general solution is
\begin{equation}
\phi ^{*}=c_{1}\,\phi _{1}^{*}+c_{2}\,\phi _{2}^{*}\,,
\label{high-energy-soln}
\end{equation}
with
\begin{eqnarray}
\phi _{1}^{*} &=&\sqrt{x}\,J_{1/6}(x),  \nonumber \\
\phi _{2}^{*} &=&\sqrt{x}\,J_{-1/6}(x)  \label{positivex}
\end{eqnarray}
(for $x>0$), where $J_{n}$ are Bessel functions of order $n$ (see
appendix \ref{BesselApp}). The coefficients $c_{1}\,,c_{2}$ will
be determined below.

In order to extend the solution (\ref{high-energy-soln}) across
$x=0$, we first express it in terms of $r$. Recall that at small
$r$,
$$
r^*\cong {\frac{r^{3}}{3\,Q^{2}},}
$$
and, furthermore, $r^*(r)$ is analytic at $r=0$. From the local
behavior of the Bessel functions $J_{n}(\lambda )$ near $\lambda
=0$ (see appendix \ref{BesselApp}), and that of $r^*$ near $r=0$,
it follows that $\phi _{1}^{*} $ and $\phi _{2}^{*}$ take the form
\begin{eqnarray}
\phi _{1}^{*} &=&r^{2}\,A_{1/6}^{J}(r^{6}),  \nonumber \\
\phi _{2}^{*} &=&rA_{-1/6}^{J}(r^{6}),  \label{fir}
\end{eqnarray}
where $A_{\pm 1/6}^{J}(\lambda )$ are some real functions (for
real $\lambda $), analytic at $\lambda =0$. Obviously both $\phi
_{1}^{*}(r)$ and $\phi _{2}^{*}(r)$ have a unique real analytic
extension through $r=0$.\footnote{Recall that $\phi _{1}^{*}(r)$
and $\phi _{2}^{*}(r)$ must extend to $r>0$ as real and analytic
functions, because the radial equation (\ref {separation-of-var})
is real and analytic (and $\phi _{1}^{*},\phi _{2}^{*}$ are real
at $r<0$).} The form of Eq. (\ref{fir}) ensures that $\phi
_{1}^{*}$ extends as an even function, and $\phi _{2}^{*}$ as an
odd one. Hence for negative $x$ we get
\begin{eqnarray}
\phi _{1}^{*}(x) &=&\sqrt{|x|}\,J_{1/6}(|x|),  \nonumber \\
\phi _{2}^{*}(x) &=&-\sqrt{|x|}\,J_{-1/6}(|x|).  \label{negativex}
\end{eqnarray}
In order to get the coefficients $c_{1}\,,c_{2}$  --  and the
transmission and reflection coefficients  --  we impose the
large-$r^{*}$ boundary conditions. Since we are considering here a
wave propagating from (say) past null infinity (large negative
$r^*$), we assume that no waves are entering from the inner
horizon $(r^{*}\to +\infty )$. This corresponds to pure asymptotic
behavior $\propto {\rm exp}(-i\,\omega \,r^{*})={\rm exp}(i\,x)$
at $x\to -\infty $. Thus, in terms of $x$ the asymptotic behavior
at the boundaries is
\begin{equation}
\begin{array}{cclcc}
\phi ^{*} & \cong  & {\rm exp}(i\,x)+R(\omega )\,{\rm exp}(-ix) &
\mbox{  at
} & x\to +\infty \,,\nonumber \\
\phi ^{*} & \cong  & T(\omega )\,{\rm exp}(ix) & \mbox{  at  } &
x\to - \infty \,.
\end{array}
\label{boundary}
\end{equation}
Note that since $\omega$ got scaled out of the equation
$T(\omega),\, R(\omega)$ are actually independent of $\omega$.

The asymptotic form of the Bessel functions for large (positive)
argument is given in Eq. (\ref{BesselAsymp}). For $n=1/6$ it reads
$$
{\rm exp}(\pm i|x|)\cong \sqrt{\pi \,|x|/2}\,{\rm exp}(\pm i\pi
/3)\left[ (1 \pm i\sqrt{3})J_{1/6}(|x|)\mp 2iJ_{-1/6}(|x|)\right]
.
$$
Applying it first to the inner-horizon boundary ($x\to -\infty $),
along with Eq. (\ref{negativex}), one finds
\begin{eqnarray*}
c_{1} &=&\sqrt{\pi /2}\,{\rm exp}(-i\pi /3)(1-i\sqrt{3})T(\omega ), \\
c_{2} &=&-i\sqrt{2\pi }\,{\rm exp}(-i\pi /3)T(\omega ).
\end{eqnarray*}
Similarly, when applied to the null-infinity boundary ($x\to
+\infty $), along with Eq. (\ref{positivex}), it yields

\begin{eqnarray*}
c_{1} &=&\sqrt{\pi \,/2}\left[ {\rm exp}(i\pi
/3)(1+i\sqrt{3})+{\rm exp}(-i \pi /3)(1-i\sqrt{3})R(\omega )\right] , \\
c_{2} &=&-i\sqrt{2\pi }\left[ {\rm exp}(i\pi /3)-{\rm exp}(-i\pi
/3)R(\omega )\right] .
\end{eqnarray*}
Combining the two expressions for $c_{1}\,,c_{2}$ and solving for
$R$ and $T$, one obtains \be
 R=\frac{\sqrt{3}}{2}i\, , \qquad T=\frac{1}{2} \,.
 \ee

We conclude that for any $l$, at the high-energy limit a fraction
$|R|^{2}=3/4$ of the influx is reflected and a fraction
$|T|^{2}=1/4$ is transmitted to the other side of the $r=0$
singularity. As was mentioned above, this applies to both waves
coming from past null infinity (of the negative-$r$ universe
beyond the singularity) and waves coming from the inner horizon
towards the $r=0$ singularity.

\subsection{Summing over $l$}

\label{summing-l}

The total cross section for transmission is given by a sum over
all partial waves $l$:
\begin{equation}
\sigma ={\frac{\pi }{k^{2}}}\sum_{l=0}^{\infty
}(2\,l+1)\,|T_{l}|^{2}, \label{total-cross-section}
\end{equation}
where $k$ is the momentum of the incoming plane wave ($k=\omega $
for a massless field, which we assume from now on in this
subsection). We shall consider here the cross section at the
high-energy limit. As mentioned above, at this limit the total
cross section is dominated by the contribution from modes $l\gg
1$, because each particular $l$ is suppressed by the factor
$\omega ^{-2}$, while it will turn out that $\sigma \gg \omega
^{-2}$.
 The result of the previous subsection, $|T_{l}|^{2}\cong 1/4$,
holds for $l$ values that are not too large. However, for any
fixed and large $\omega $ there is some large enough $l$ where the
transmission practically vanishes. For instance, for classical
flat-space scattering off a rigid target of length scale $L$
(``rigid'' means here $\omega $-independent $L$), if the impact
parameter $b$ is much larger than $L $ then the absorption
vanishes. Since $b=l/p$ (and $p=\omega $ in our massless case) we
see that for $l>L\omega $ the absorption vanishes. We will now
show that in our case the suppression of transmission already
happens for $l>(Q\omega \,)^{1/3}$.

A rough estimate of the total large-$\omega$ cross section can be
easily done based on the qualitative features of $V_{{\rm eff}}$
at small $r$. For $\omega \gg \omega _{peak}$, where $\omega
_{peak}$ is defined in (\ref{wpeak}), the above high-energy result
$|T_{l}|^{2}\cong 1/4 $ holds. For $\omega $ significantly smaller
than $\omega _{peak}$ the incoming wave will encounter the
potential barrier \be
 V_{{\rm eff}}\cong \frac{l^{2}}{Q^{2/3}(3r^*)^{4/3}}
 \label{Vbarrier}
  \ee and will be tunnelling suppressed there. [In fact, an
effective tunnelling suppression requires that $\omega \Delta$ is
greater than $1$, where $\Delta $ is the tunnelling length scale
in the barrier in terms of $r^*$, i.e. from (\ref{Vbarrier})
$\Delta \sim (\omega ^{3}Q/l^{3})^{-1/2}$. But this is
automatically satisfied because $\omega \Delta \sim (\omega
Q/l^{3})^{-1/2}$ is roughly $(\omega _{peak}/\omega )^{1/2}$,
which is $\gg 1$ in the present discussion.] Thus, for a rough
estimate of the large-$\omega $ cross section we may substitute
$|T_{l}|^{2}\cong 1/4$ for $l\ll l_{c}$ and negligible
$|T_{l}|^{2}$ for $l\gg l_{c}$, where \be
 l_{c}:= \sqrt{3}(Q\omega )^{1/3} \ee
can be read off the expression for $\omega_{peak}$ (\ref{wpeak}).

Equation (\ref{total-cross-section}) then yields the total cross
section:
\begin{equation}
\sigma \sim \frac{l_{c}^{2}}{\omega ^{2}}\sim
{\frac{Q^{2/3}}{\omega ^{4/3}}.}  \label{cross-section1}
\end{equation}
The total cross section may also be obtained from $l_c$ by a simple argument:
the critical maximal impact parameter for
transmission is $b_{\rm max} = l_c/ \omega$, and hence the total size
(cross-section) of the target for transmission is
\begin{equation}
\sigma \sim b_{\rm max}^2 = \frac{l_{c}^{2}}{\omega ^{2}}\sim
{\frac{Q^{2/3}}{\omega ^{4/3}}. }  \label{cross-section2}
\end{equation}

In the above estimate we have not taken into account the
contribution coming from the range $l\sim l_{c}$. A more rigorous
derivation of the cross section
(\ref{cross-section1},\ref{cross-section2}), which properly treats
this range as well, may be done by analyzing the rescaling
properties of the small-$r$ scattering problem, i.e.
\begin{equation}
\lbrack -{\partial }_{r^{*}\,r^{*}}+V_{{\rm {eff}}}-\omega
^{2}]\phi =0
\end{equation}
with the potential (\ref{Vsmallr*}). Transforming to a new
dimensionless variable
\begin{equation}
x:=-{\frac{l^{3}}{\,Q}r^{*},}
\end{equation}
the equation becomes
$$
\lbrack -{\partial }_{x\,x}+\hat{V}(x)-\Omega ^{2}]\phi =0,
$$
where
$$
\hat{V}(x)=-{\frac{2}{9\,x^{2}}}+{\frac{1}{(3x)^{4/3}}}
$$
and
$$
\Omega :={\frac{Q}{l^{3}}\omega .}
$$
The boundary conditions at large $|x|$ are, in accordance with Eq.
(\ref{boundary}),
\begin{equation}
\begin{array}{cclcc}
\phi ^{*} & \cong  & {\rm exp}(i \Omega x)+R(\Omega )\,{\rm exp}(-i \Omega x)
& \mbox{  at } & x\to +\infty \,,\nonumber \\
\phi ^{*} & \cong  & T(\Omega )\,{\rm exp}(i \Omega x)
& \mbox{  at  } & x\to -\infty \,.
\end{array}
\label{boundary1}
\end{equation}
Hence in this limit of large $\omega $ and large $l$ the
transmission amplitude depends only on $\Omega$, i.e.
\begin{equation}
T=T({\frac{\omega \,Q}{l^{3}}}).
\end{equation}
The cross section is
$$
\sigma ={\frac{\pi }{\omega ^{2}}}\sum_{l=0}^{\infty
}(2\,l+1)\,|T_{l}|^{2} \cong {\frac{2\pi }{\omega ^{2}}}\int
\,l\,\,|T({\frac{\omega \,Q}{l^{3}}} )|^{2}dl,
$$
where we used $l\gg 1$ to replace $2\,l+1$ by $2\,l$ and the sum
by an integral. Finally, transforming from $l$ to
$\hat{l}:=l/(\omega \,Q)^{1/3}$ we find
\begin{equation}
\sigma \cong c{\frac{Q^{2/3}}{\omega ^{4/3}},}
\end{equation}
where
$$
c=2\pi \int \,\hat{l}\,\,|T(\hat{l}^{-3})|^{2}d\hat{l},
$$
in agreement with the above estimate (\ref{cross-section1}).

Having found the total cross section for transmission, one may
inquire about the total cross section for scattering. At large
distances this is just like a Coulomb problem with a length scale
$\,M$. Due to the infinite range of the gravitational force, we
expect the cross section for scattering to be infinite
for a massive field.
However, for a massless field the scattering potential decays
(after subtracting the standard centrifugal piece
$l(l+1)/r^{*~2}$) as $\ln (r^*/M)/r^{*~3}$. This is a short-range
potential, hence we expect a finite scattering cross-section in
the massless case.

The peculiar $\omega$ dependence of the transmission cross section
(\ref{cross-section1}) through the singularity can be compared
with two other systems. For black holes with non-zero area the
large $\omega$ absorption cross section through the event horizon
tends to the geodesic cross section, and hence $\sigma \sim
\omega^0$. For elementary particles, on the other hand, the
high-energy scattering cross section $\sigma_{\rm scat} \sim
\omega^{-2}$ which is the kinematical factor which multiplies the
dimensionless scattering amplitude. So we see that this system
lies between the two examples above.

Another interpretation issue is whether the partial
transmission/partial scattering behavior at large $\omega$ should
be interpreted as a beam splitter for single particles. The
question is whether a wave packet can be made small enough so that
it fits into a target of size $\sim \omega^{-2/3}$ -- otherwise it
will not be fully split. Imagine that we set a ``lens'' (one which
affects $\phi$ waves) at a distance $f \sim \omega^{0} \gg Q$ from
the black hole. We know that the spot size at focus can be made as
small as $\lambda\, f/D$, where $\lambda=2\, \pi/\omega$ is the
wavelength and $D$ is the aperture size, and we here imagine that
space is flat and we need to hit a target with the same cross
section as the $r=0$ singularity. From $\lambda\, f/D \ll
\omega^{-2/3}$ we find that we need $D \ge \omega^{-1/3}$. So such
focusing can be readily achieved, even with a small size lens.
This suggest that the singularity may indeed act as a beam
splitter. Note, however, that in the above analysis we only
considered incoming {\it plane waves} (or pure partial waves with
well-defined $l$ values). In order to verify the beam-splitter
behavior one needs to explicitly analyze the transmission
amplitude in the geometrical configuration considered above (i.e.
a narrow beem focused on a small region near $r=0$). This
calculation is beyond the scope of the present paper.

\section{Generalizing the backgrounds}
\label{generalizations}

\subsection{Higher dimensions}
\label{highd}

The background RN metric and gauge field in $d \ge 4$ dimensions
are \cite{MyersPerry} \bea
 ds^2 &=& -f\, dt^2 + f^{-1}\, dr^2 + r^2\, d\Omega_{d-2}^{~2} \non
 A &=& -{Q \over r^{d-3}}\, dt \non
 f &=& 1-{r_+^{d-3}+r_-^{d-3} \over r^{d-3}} + {(r_+\, r_-)^{d-3} \over
r^{2\,(d-3)}}
  \label{RNmetric2}
 \eea
 The mass and charge are \bea
 M &=& {(d-2)\, \Omega_{d-2} \over 16\, \pi}( r_+^{d-3} + r_-^{d-3} ) \non
 Q &=& {\rm const}\,\sqrt{r_+ \, r_-}^{~d-3} \eea
where $\Omega_{d-2}$ is the area of the $d-2$ sphere, and we do
not fix the constant in front of the charge which depends on the
normalization chosen for the vector field. In order to get
$d$-dimensional Schwarzschild one sets $r_-=0, ~ r_+=r_0$.

After separation of the time variable ($\del_{tt} \rightarrow
-\omega^2$) the scalar wave equation becomes \be
  \br  {1 \over f\, r^{d-2}}\, \del_r\, f\, r^{d-2}\, \del_r -
 {l\, (l+d-3) \over f\, r^2} + f^{-2}\, \omega^2 \kt \,
 \phiom =0 ~.
 \label{SeparatedWaveEqd} \ee

Considering the charged case first, and comparing with the 4d case
(\ref{rnreg}) one notes a change: the equation
(\ref{SeparatedWaveEqd}) is not regular anymore at $r=0$ (since
$f\, r^{d-2}$ is has a pole).
 Actually it is regular-singular with leading behavior \be
  \br -\del_{rr} + {d-4 \over r}\, \( 1+O \( r^{d-3} \) \)
  \del_r + l\, (l+d-3)\, O \( r^{2(d-4)}\) + \omega^2\, O \( r^{4(d-3)} \) \kt\, \phiom =0~.
  \label{reg-sing-dg4} \ee
The two solutions of the characteristic equation are $r^0,\,
r^{d-3}$. Since the difference of the exponents is integral it is
possible {\it a priori} that the $r^0$ solution has a $\log$ piece
of the following form $\phiom=r^0 + \dots+ \log(r)\, r^{d-3} +
\dots $, however it is seen not to be the case by explicitly
Taylor expanding $\phiom$ in the differential equation
(\ref{reg-sing-dg4}).

For Schwarzschild, on the other hand, one finds the same leading
behavior as in 4d \be
 \br \del_{rr} + {1 \over r}\,  \del_r + \dots \kt \phiom=0 \ee
and thus the leading behavior of the two independent solutions is
still $r^0,\, \log(r)$.

As in subsection \ref{normal-form-subsection} we may find the
normal form of the radial equation (\ref{SeparatedWaveEqd}). The
$r^*$ coordinate is defined by \be
 dr^* = {dr \over f} ~~,\ee
 the field is redefined by \be
 \phi^* := r^{(d-2)/2}\phi~, \ee
and the resulting effective potential is
 \bea V_{\rm eff} &=& \Delta V_{\rm eff} + f\, {l\, (l+d-3) \over r^2} \non
   \Delta V_{\rm eff} &=& {d-2 \over 2}\, {f \over r^{(d-2)/2}} \del_r
(r^{(d-4)/2}\, f)
  \label{Veffd} \eea
where we use the same notations as in subsection \ref{normal-form-subsection}.

We start by looking at the charged case. The leading behavior near
the singularity (at $r=0$) is \bea
 r^* &\cong& {r^{2\, d-5} \over (2 \, d -5)\, (r_+\, r_-)^{(d-3)}}  \non
 V_{\rm eff} &\cong& -{(d-2)\, (3d-8) \over 4\, (2d-5)^2}\, {1
\over r^{*~2} }   \label{Veff-sing-RN} \eea
 Given an effective potential $V_{\rm eff} \simeq -c/r^{*~2}$ the order
$n$ of the Bessel functions which appear in the high $\omega$
solution is given by (see appendix \ref{BesselApp}) \be
 n^2 = {1\ \over 4} - c ~~.\ee
Hence, the second equation in (\ref{Veff-sing-RN}) determines \be
 |n|_{\rm RN}={1 \over 4} \, {d-3 \over  d-5/2} \ee
 generalizing the 4d result $|n|=1/6$ (\ref{positivex}).

We note in passing that for the 2d black hole we will show
in the next subsection
that $V_{\rm eff} \simeq -3/(16 r^{*~2})$ and hence one can define
an ``effective'' or ``equivalent'' RN dimension to be $d_{\rm
eff}= \infty$ or $11/4$, and $|n|_{2d}=1/4$.

The transmission and reflection coefficients for high $\omega$
 and fixed $l$ follow from the asymptotics of the Bessel functions as in
subsection \ref{trasnmission-reflection}:
 \bea |R| &=&  |\cos (n\, \pi)| \non
     |T| &=&    |\sin (n\, \pi)| \label{rrtt} \eea
The $\omega$ dependence of the total cross section for
transmission can be derived along the lines of subsection
\ref{summing-l}. The leading terms in the potential (\ref{Veffd})
at $r \sim 0$ are \be
 V_{\rm eff} \simeq l^2\, {(r_+\, r_-)^{d-3} \over r^{2\, d-4}}-
 {d-2 \over 2}\, ({3 \over 2}\, d -4)\, {(r_+\, r_-)^{2(d-3)} \over r^{4\,
 d-10}} ~ .\ee
 For fixed $l$ the potential has a maximum at \be
 r_{peak} \simeq {1 \over \sqrt[d-3]{l}} ~.\ee
 Equating $V(r_{peak})$ with $\omega^2$ we get the transition value
 for $l$: \be
 l_c \sim \omega^{d-3 \over 2\, d -5}~. \ee
Finally, we evaluate the total transmission cross section by
employing qualitative classical-mechanics considerations.
Classically the cross section is obtained by estimating the
maximal value of impact parameter $b$ for which a significant
transmission (or scattering, in the more general context) still
occurs. Using $b=l/\omega$ we get $b_{\rm max} \sim l_c/\omega$.
Since in $d$ dimensions the cross section scales as $b_{\rm
max}^{d-2}$, we find \be
 \sigma \sim b_{\rm max}^{d-2} \sim \left( {l_c \over \omega}
 \right)^{d-2} \sim \omega^{-{(d-2)^2 \over 2\, d-5}}~. \ee

For \Schw black holes the leading behavior near the singularity
(at $r=0$) is \bea
 r^* &\cong& -{r^{d-2} \over (d -2)\, r_0^{d-3}}  \non
V_{\rm eff} &\cong& -{1 \over 4\, r^{*~2} }
    \label{Veff-sing-Schw} ~.\eea
Thus independently of dimension one has $c=1/4$ and hence total
reflection: \bea
 n &=& 0 \non
 T &=& 0 ~~.\eea

\subsection{The 2d black hole}
\label{tdss}

Black holes in two dimensions are apparently outside the pattern
of the $d\geq 4$ black holes discussed in the previous sections
(although as we shall recall \cite{GiveonRabinoviciSever} they
turn out to share similar features). Such solutions require, for
instance, the presence of a dilaton. In this subsection we shall
consider two dimensional dilaton gravity with an Abelian gauge
field which is inspired by string theory. The action
is~\footnote{This can be obtained, for instance, as (part of) a
low-energy effective action of an heterotic string in two
dimensions (for a review, see \cite{Polchinski}).}
 \be
   S=\int d^2 x\sqrt{-g}e^{-2\Phi}\left(
   R+4g^{\mu\nu}\del_\mu\Phi\del_\nu\Phi-{1\over 4} F^2-\lambda\right)~,
 \ee
where $g_{\mu\nu}$ is the (string frame) metric, $\mu,\nu=0,1$,
$\Phi$ is the dilaton, $F_{\mu\nu}$ is the field strength of an
Abelian gauge field $A_{\mu}$, and $\lambda$ is the cosmological
constant. In this two dimensional theory, the charged black hole
solution is given by \cite{nappiyost}
 \bea
    ds^2 &\propto & f^{-1}\, d\rho^2 -f\, dt^2 \non
    A &\propto& -{Q\over r}\, dt \non
    \Phi(\rho) &=& \Phi_0- \half \rho~, \label{twodsol}
 \eea
where $\Phi_0$ is a constant and
 \be
    f(r)=1-{2M\over r}+{Q^2\over r^2}\ ,\qquad r=e^\rho~.
    \label{twodf}
 \ee
Note that here $ds^2\propto f^{-1}\, {dr^2\over r^2} -f\, dt^2$,
which is different from the generic $d\geq 4$ cases in eq.
(\ref{RNmetric2}). Nevertheless, the geometry of this 2d
black-hole is similar to a two dimensional slice of the
Reissner-Nordstr\"{o}m solution, whose complete Penrose diagram is
given in figure \ref{PenroseRN}. In the 4d case every point in
figure \ref{PenroseRN} is actually a two sphere, while in the 2d
case there is a non-trivial dilaton instead.

An uncharged scalar field is minimally coupled to the background
above as follows: \be
   S=\int d^2 x\sqrt{-g}e^{-2\Phi}\left(
   R+4g^{\mu\nu}\del_\mu\Phi\del_\nu\Phi-{1\over 4} F^2-\lambda -
   g^{\mu\nu}\del_\mu \psi\del_\nu \psi-m^2\psi^2\right)~,
 \ee
where in this section
$\psi$ denotes the scalar field to distinguish it
from the dilaton $\Phi$. The wave equation of $\psi$ is:
 \be
    \Box_\Phi\psi={e^{2\Phi}\over\sqrt{-g}}\, \del_\mu\,
    \sqrt{-g}\, e^{-2\Phi}\, g^{\mu\nu}\,
    \del_\nu\psi=\(\del_rr^2f\del_r-f^{-1}\del_{tt}\)\psi=
    m^2\psi~, \label{laplace}
 \ee
where we used $g_{rr}=1/(f\, r^2)$ and $\sqrt{-g}=e^{2(\Phi -\Phi_0)}=1/r$.
Hence, in the two dimensional case the dilaton $\Phi$ plays again
the role of the 4d spherical coordinates, now in the wave
equation.

As in the $d\geq 4$ cases, we separate variables
 \be
    \psi(r,t)=\psi_\omega(r)e^{i\omega t}~.
 \ee
The equation for $\psi_\omega$ is:
 \be
    \(     \del_rr^2f\del_r+\omega^2f^{-1}-m^2\)\psi_\omega = 0~.
    \label{eqpsi}
 \ee
Again, we define a coordinate $r^*$,
 \be
    dr^*={dr\over rf(r)}~,
 \ee
and a new field variable
 \be
    \psi^{*}:= \sqrt{r}\psi~,
 \ee
such that eq. (\ref{eqpsi}) turns into:
 \be
    \left[-\del_{r^*r^*}+V_{{\rm eff}}(r^*)-\omega^2\right]\psi^{*}_\omega=0~.
 \label{turnsin}
 \ee
The effective potential is now
 \be
    V_{{\rm eff}}(r^*)=\Delta V_{{\rm eff}}+fm^2~,
 \ee
where
 \be
    \Delta V_{{\rm eff}}=\half\sqrt{r}f\del_r(\sqrt{r}f)~.
 \ee
Next we inspect the behavior of the wave equation (\ref{eqpsi})
near the singularity of the 2d RN-like black hole. When $Q\neq 0$:
 \bea
    &{1\over Q^2}\(\del_r r^2f\del_r+\omega^2f^{-1}-m^2\)\psi_\omega=
    \non &\left[(1+...)\del_{rr} +O(r^0)\del_r-{m^2\over Q^2}+
    (1+...){r^2\over Q^4}w^2\right]\psi_\omega = 0~,
 \eea
where ``$...$'' stand for higher orders in $r$. As in the 4d case
(\ref{rnreg}), this equation is regular at $r=0$ for every
$\omega$. Moreover, near $r\sim 0$ we have
 \be
    r^*\cong{r^2\over 2Q^2}
 \ee
and hence
 \be
    V_{{\rm eff}}(r^*)\cong -{3\over 16\, r^{*~2}}\label{veff}~.
 \ee
On the other hand, for $Q=0$ -- the Schwarzschild-like 2d black
hole -- the equation for $\psi_\omega$ is:
 \bea
    &-{1\over 2Mr}\(\del_rr^2f\del_r+\omega^2f^{-1}-m^2\)\psi_\omega=
    \non &\left[(1+...)\del_{rr} +{(1+...)\over r}\del_r+{m^2\over 2Mr}+
    (1+...){\omega^2\over 4M^2}\right]\psi_\omega = 0~.
 \eea
Again, as in the 4d case (\ref{Schw-var-sep}), it is not regular
at the singularity, but instead it is regular-singular. In the
uncharged case, to leading order near $r\sim 0$ we have
 \be
    r^*\cong -{r\over 2M}
 \ee
and hence
 \be
    V_{\rm eff}(r^*) \cong -{1 \over 4\, r^{*~2}}~.\label{veffunch}
 \ee
As discussed in previous sections, the solutions to high frequency
scattering wave equations (\ref{turnsin}) are given in terms of
Bessel functions, which depend on the effective potential. In the
charged case we read the order of the relevant Bessel function
from (\ref{veff}) and appendix \ref{BesselApp}: \be |n|={1 \over
4}~, \ee and hence at high energies we find (\ref{rrtt}):
 \be
    |T| = |R| = {1 \over \sqrt{2}}~. \label{rt}
 \ee
On the other hand, for the uncharged case, from (\ref{veffunch})
and (\ref{rrtt}) we read:
 \bea
    n &=& 0 \non
    |R| = 1&,& \quad T = 0~. \label{rtunch}
 \eea
To summarize, we see that the features of the 2d black holes
discussed in this subsection are very similar to the 4d case.

The 2d background (\ref{twodsol}) is also an exact Conformal Field
Theory (CFT) background in string theory; in
\cite{GiveonRabinoviciSever} it was obtained from a family of
${SL(2,\IR)\times U(1)\over U(1)}$ quotient CFT sigma models by a
Kaluza-Klein reduction\footnote{In the bosonic case it is a
semi-classical approximation, while in the superconformal
extension this background is claimed to be exact \cite{bars}.}. In
string theory one is forced to include the regions beyond the
singularity \cite{giveon} (this is argued, for instance, by using
T-duality; for a review, see \cite{gpr}). The structure of the
parent $SL(2,\IR)$ group allows one \cite{GiveonRabinoviciSever}
to find the exact solutions to the wave
equation\footnote{Scattering waves are determined from vertex
operators in the $SL(2,\IR)$ CFT. Those are given by matrix
elements in a unitary representation of the group. Equivalently,
these are solutions to the Laplace equation (\ref{laplace}), which
in this case is a solvable hypergeometric equation.}. In
particular, it gives the exact reflection coefficient
 \be
    |R(\omega)|^2={\ch\({r_-\omega \over r_+-r_-}\)\ch(\omega)
\over\ch\({r_+\omega \over r_+-r_-}\)}~,
 \ee
where $\omega$ is proportional to the energy of a massless
particle scattered from the region beyond the black hole
singularity, and $r_{\pm}$ are the locations of the event and
Cauchy horizons:
 \be
   r_\pm=M\pm\sqrt{M^2-Q^2}~.
 \ee
Indeed, for the charged case $|R|^2 \longrightarrow \half$ in the
limit $\omega\rightarrow\infty$, and for the uncharged case
$|R|^2=1$ for every energy. These results are in agreement with
(\ref{rt}) and (\ref{rtunch}), respectively.

\section{Adding perturbations}
 \label{perturbations-section}

Until now we assumed that $\phi$ was a scalar field with no mass,
no charge, no back-reaction and no other interactions beyond
minimal coupling to gravity. Now we will check whether our result
that the time evolution is well-defined is disrupted  by any of
these perturbations. To our surprise we do not encounter any
problem.

\subsection{Mass and charge}

Adding a mass $m$ and charge $q$ for the field $\phi$ is
completely captured by the form of the geodesic potential
(\ref{Veff-geod}), namely one has \bea
 V_{\rm eff}^{\rm wave} &=& \tilde{V}_{\rm eff}^{\rm geodesic} + \Delta
V_{\rm eff} \non
 \tilde{V}_{\rm eff}^{\rm geodesic} &=&
\left({\frac{l\,(l+d-3)}{r^{2}}}\,+m^{2}\right)\,f(r)
  -\left(E-{\frac{q\,Q}{r^{d-3}}}\right)^{2} \non
 \Delta V_{\rm eff} &=& {d-2 \over 2}\, {f \over r^{(d-2)/2}} \del_r
(r^{(d-4)/2}\, f)
 \eea
 where $\tilde{V}_{\rm eff}^{\rm geodesic}$ is simply
 $V_{\rm eff}^{\rm geodesic}$ from (\ref{Veff-geod}) with the
 substitutions
 $l^2 \to l(l+d-3)$
and $E \to \omega$,
and $\Delta V_{\rm eff}$ is the same as in the zero-mass
zero-charge case (\ref{DeltaVeff},\ref{Veffd}).

Our analysis relied on the leading terms near $r \sim 0$, which we
will now find to dominate over the added perturbations. For $Q
\neq 0$ these are $1/r^{4 d -10}$ from $\Delta V_{\rm eff}$ and
$l^2/r^{2d-4}$ from $\tilde{V}_{\rm eff}^{\rm geodesic}$. Since
both $m$ and $q$ contribute at order $1/r^{2d-6}$, they are
irrelevant near the singularity (for $d \ge 4$). Moreover, one can
confirm that the form of the subleading terms still conform with
(\ref{reg-sing-dg4}) and hence there are no log pieces in the
solutions.

Note however that since for $q \neq 0$ one cannot solve for
$\omega^2$ in the wave equation and view it as eigen-values of
some operator, any operator approach and considerations of
self-adjointness (see section \ref{conditions-section}) would need
to be changed.

Finally, for $Q=0$ we might as well take $q=0$ and the leading
terms are $1/r^{2 d -4}$ from $\Delta V_{\rm eff}$ and
$l^2/r^{d-1}$ from $\tilde{V}_{\rm eff}^{\rm geodesic}$, while $m$
contributes at $1/r^{d-3}$ and is again irrelevant.

\subsection{Interactions}

We may add non-quadratic terms to the action, namely adding terms
non-linear in $\phi$ to the wave equation. Much of our previous
analysis, especially the separation of variables relied on
linearity, hence we should reconsider it, starting from the
non-linear wave equation \be
 0=\Box \phi - g\, U(\phi) ~, \ee
 where $g$ is a coupling constant and $U(\phi)$ is a non-linear
potential term. One can attempt to solve this equation by
perturbation theory with $g$ being the small parameter, namely one
expands \be
 \phi = \sum_j g^j\, \phi^{(j)} ~.\ee
We start at zeroth order with a solution $\phi^{(0)}$ to the
linearized equation \be
 \Box \phi^{(0)}=0 \ee
and consider whether the first correction is regular at $r=0$. The
first correction satisfies \be
 \Box \phi^{(1)} = g\, U(\phi^{(0)}) ~. \label{interactions-first-order} \ee
 In order to solve this equation we separate the angular and time
 variables \be
 f^{-1}\, \Box_{\omega l}\phi^{(1)}_{\omega l} = g\, f^{-1}\, \left[
U(\phi^{(0)})
 \right]_{\omega l} ~. \ee
 Since $\phi^{(0)}$ is regular by assumption, so are the
components  $\left[ U(\phi^{(0)})
 \right]_{\omega l}$. This is a non-homogeneous ordinary (second order)
differential equation. A solution in the vicinity of the
singularity may be obtained by Taylor expanding the equation (see
also section \ref{conditions-section}). The homogeneous solutions
are the smooth $\phi^{(0)}_{\omega l}$,
 while the non-homogenous term added to (\ref{SeparatedWaveEqd}) is
$O\(f^{-1}(r)\)$ and hence it is subleading and does not change
our ability to obtain a (univalued) solution. This will happen at
higher orders in $g$ as well.

\subsection{Back-reaction for RN in $d \ge 4$}

So far we worked in the linear approximation where one neglects
the back-reaction of the scalar field on the background geometry.
Let us check whether including the back-reaction disrupts our
result, namely whether we can find any obstruction for the
solutions to the linear equation from being extended to solutions
for the full non-linear system of background plus scalar field.
Our analysis will be limited to lowest order back-reaction, and we
shall not cover all cases, but the indications are that these
solutions do survive back-reaction.

Let us define a small parameter $\eps$ such that $\phi$ is first
order in $\eps$. Then at second order the background may change --
the metric due to a $T_{\mu\nu}$ source, and the electromagnetic
(EM) field due to the charge associated with $\phi$. For
simplicity we may consider $\phi$ to be neutral. Since the metric
is singular to start with, we cannot use the straight-forward
criterion that the background remains regular after back-reaction.
A conservative view would be to continue and look at third order
whether the changes in $\phi$ as a result of the changes in the
background make it singular. Another approach is to require that
the singularity in the metric does not become worse after
back-reaction. The latter has the advantage that one can stop at
second order.

Moreover, for a charged black hole we can even save us the work of
computing the second order by the following observation. Let us
compare the stress-energy tensor of the scalar field with that of
the EM field, namely the background source.
The action for gravity + EM + an uncharged scalar field is
 \be
   S=\int d^d x\sqrt{-g} \left(
   R-{1\over 4}F^2- g^{\mu\nu}\del_\mu \phi\del_\nu
   \phi-m^2\phi^2 \right)~,
 \ee
where $g_{\mu\nu}$ is the metric, $F$ the field strength and
$\phi$ an uncharged scalar field of mass $m$. The stress tensors
of the gauge and scalar fields are
 \bea
   T_{\mu\nu}^{(F)}&=&\half F_{\mu\rho}F_{\nu\sigma}g^{\rho\sigma}-{1\over
   8}g_{\mu\nu}F^2 \non
   T_{\mu\nu}^{(\phi)}&=&\del_\mu\phi\del_\nu\phi- \half
   g_{\mu\nu}\left[(\del\phi)^2+m^2\phi^2\right] \label{stresst}
 \eea
We start by comparing the scalar quantities \bea
 (\del \phi)^2 &=& \del_r \phi\, \del_r \phi\ g^{rr} \sim  {1 \over
r^{2(d-3)}} \non
 (F^2)  &=& F_{rt}\, F_{rt}\, g^{rr}\, g^{tt} \sim {1 \over
 r^{2(d-2)}} \eea
 where we use the background (\ref{RNmetric2}) and the property that
$\phi,\, \del_r\phi$ are regular at the singularity. We see that
close to the singularity the scalar field seems to add a
negligible source on top of the one from the EM field.

We now proceed to compare all components
 \be \begin{array}{ll}
 T_{rr}^{(F)} \sim  {1\over r^2}    \qquad     &    T_{rr}^{(\phi)} \sim 1 \\
 T_{tt}^{(F)} \sim  {1\over r^{2(2d-5)}} \qquad &    T_{tt}^{(\phi)} \sim
{1\over r^{4(d-3)}} \\
 T_{rt}^{(F)}=0                          \qquad  &    T_{rt}^{(\phi)} \sim 1 \\
 T_{\theta \theta}^{(F)} \sim  {1\over r^{2(d-3)}} \qquad &    T_{\theta
\theta}^{(\phi)} \sim {1\over r^{2(d-4)}} \\
\end{array}
 \ee
We see that indeed for the $rr,~tt$ and $\theta \theta$ components
$T_{\mu \nu}^{(\phi)} \sim r^2 \, T_{\mu \nu}^{(F)}$ is
negligible, while $T_{rt}^{(\phi)}$ is regular. Altogether we
consider this to be strong evidence that back-reaction is weak and
does not change the smoothness of the solutions.

For a charged field $\phi$ the simple argument above does not seem
to work. In principle one should determine the back-reaction to
the metric and gauge field and whether they are more or less
singular than the original background. A preliminary analysis
turned out to be involved, and so we did not reach any conclusions
for this case. For similar reasons we do not discuss the
back-reaction to the negative-mass \Schw either.

\subsection{Back-reaction for the 2d black-hole}
In two dimensions, the action for an uncharged scalar field
coupled to the metric and dilaton is (see subsection \ref{tdss}):
 \be
   S=\int d^2 x\sqrt{-g}e^{-2\Phi}\(
   R+4g^{\mu\nu}\del_\mu\Phi\del_\nu\Phi-{1\over 4} F^2-\lambda -
   g^{\mu\nu}\del_\mu \psi\del_\nu \psi-m^2\psi^2\right)~,
 \ee
where $g_{\mu\nu}$ is the string frame metric, $\Phi$ the dilaton,
$F$ is the field strength of an Abelian gauge field $A$, $\psi$ is
the scalar field (to distinguish it from the dilaton) and
$\lambda$ is the cosmological constant.

Using the background (\ref{twodsol}) and the property that
$\psi,\, \del_r\psi$ are regular at the singularity, we find the
following behavior of the scalar stress tensor (\ref{stresst})
near the singularity
 \bea
    e^{2\Phi}T_{rr}^{(\psi)}&\sim&1 \non
    e^{2\Phi}T_{tt}^{(\psi)}&\sim&{1\over r^2} \non
    e^{2\Phi}T_{rt}^{(\psi)}&\sim&1
 \eea
The dilaton stress tensor is
 \be
   e^{2\Phi}T_{\mu\nu}^{(\Phi)}=4\nabla_\mu\nabla_\nu\Phi
   -4g_{\mu\nu}\left(\nabla^2\Phi-(\partial\Phi)^2-{1\over 4}\lambda\right)
 \ee
and has the following leading behavior near the singularity
 \bea
    e^{2\Phi}T_{rr}^{(\Phi)}&\sim&{1\over r^2} \non
    e^{2\Phi}T_{tt}^{(\Phi)}&\sim&{1\over r^4} \non
    e^{2\Phi}T_{rt}^{(\Phi)}&=0
 \eea
Again, there is no strong back-reaction to the metric, since
component-wise each component of $T_{\mu\nu}^{(\psi)}$ is either
subleading to $T_{\mu\nu}^{(\Phi)}$ or regular.

To check the back-reaction to the dilaton, we define
 \be
    \chi=e^{-\Phi} ~~,
 \ee
in terms of which the action becomes
 \be
   S=\int d^2 x\sqrt{-g}\br4g^{\mu\nu}\del_\mu\chi\del_\nu\chi+\chi^2
\(R-{1\over 4} F^2-\lambda -
   g^{\mu\nu}\del_\mu \psi\del_\nu \psi-m^2\psi^2\)\kt~.
 \ee
The equation of motion for $\chi$ is
 \be
    4\Box\chi-\(R-{1\over 4} F^2-\lambda -
    g^{\mu\nu}\del_\mu \psi\del_\nu \psi-m^2\psi^2\)\chi=0~,
    \label{eom}
 \ee
where $\Box$ is the Laplacian without the dilaton. In principle,
we should expand the solution to (\ref{eom}) to second order in
the perturbation and compare the correction to its initial
value~\footnote{At zero order we have $\chi=\chi_0\sqrt{r}$
(\ref{twodsol}, \ref{twodf}). The second solution to (\ref{eom})
is subleading near $r=0$, so turning it on at higher orders would
not change the result.}. However, just like for the metric, we
recognize that the new source for $\chi$, ~($g^{\mu\nu}\del_\mu
\psi\del_\nu \psi-m^2\psi^2$), is negligible at the singularity
relative to the background sources ($R-{1\over 4} F^2$).
Therefore, we do not expect a strong back-reaction for the dilaton
either. Summarizing, we find indications for a weak back-reaction
near the singularity of the 2d charged black-hole.


\section{General conditions for resolution of a singularity}
\label{conditions-section}

\subsection{Uniqueness of decomposition and natural boundary conditions}

In order for the ``singularity smoothing'' procedure which we
defined in section \ref{smooth-section}  to make sense we need the
existence and uniqueness of decomposition of any initial condition
into a linear combination of eigen-functions $\phiom$. We know
that Hermitian operators have this property, namely their
eigen-functions form a basis of function space. Thus we are
interested in representing the radial equation $\Box_{\omega l}\,
\phiom=0$ through an operator and studying its Hermiticity, and so
we define the operator $L$ by
 \bea
 f\, \Box_{\omega l} &=& \omega^2 - L[\del_r,r] \non
 L &:=& -{f \over r^2}\, \del_r\, f\, r^2\, \del_r +
 {f\, l\, (l+1) \over r^2}  \label{defL} \eea
and we specialize back to 4d for definiteness. We note that $L$ is
not regular at $r=0$ even for RN but only due to an overall factor
of $f^2$ relative to the regular radial equation $f^{-1}\,
\Box_{\omega l}$ (\ref{separation-of-var}). One may wonder whether
the decomposition property would nevertheless hold for $L$. This
will be answered in the negative as we shall show that the
functions $\phiom$ do not span all possible functions.

Since the differential equation (\ref{separation-of-var}) is
smooth at $r=0$ one can Taylor expand the equation and solutions
\be \phi_\omega(r) = \sum_{j=0}^{\infty}\, \phi_\omega^{(j)}\, r^j
\label{Taylor-phi} ~.\ee
 Substituting this in  the differential equation transforms it into a
recurrence equation for the Taylor coefficients
$\phi_\omega^{(j)}$.

For RN one notes that the coefficient of $\omega^2$ is $f^{-2}$
which is $O(r^4)$ and hence the recurrence equation is
$\omega$-independent up to order $O(r^3)$ (inclusive) where
$\phi_\omega^{(5)}$ is determined. Among the first 6 Taylor
coefficients $\phi_\omega^{(0)}, \dots, \phi_\omega^{(5)}$ the
first two may be considered to be initial conditions for the
recurrence equation and the other four may be considered to be a
function of them. In other words we find 4 $\omega$-independent
linear relations among $\phi_\omega^{(0)}, \dots,
\phi_\omega^{(5)}$. Owing to the $\omega$-independence of these
relations they continue to hold for any function which can be
expressed as a linear combination \be
 \phi_i= \int\, d\omega\, \, \( \phi_+(\omega) + \phi_-(\omega) \)\,
 \phi_{\omega}(r)~,
\label{linear-combination} \ee
 where $\phi_\pm(\omega)$ are arbitrary coefficients.
This means that there are 4 conditions on the first 6 coefficients
of any function in span($\phiom$).

For \Schw the phenomenon is similar but the details are different.
We have $f^{-2}= O(r^2)$ and hence the recurrence relation is free
up to (and including) the 4th coefficient $\phi_\omega^{(3)}$.
Since we consider only the regular solutions then the recurrence
relation requires a single initial condition $\phiom^{(0)}$  while
$\phi_\omega^{(1)}, \dots, \phi_\omega^{(3)}$ may be considered to
be an $\omega$-independent function of it, and hence there are
necessarily 3 conditions on the first 4 Taylor coefficients of any
function in span($\phiom$).

We see that ${\rm span}(\phiom)$ does not contain all functions,
but rather there are constraints on the initial conditions at
$r=0$. This is why we required the time-evolution to be
well-defined only for wave packets prepared far away, but
otherwise arbitrary. We expect that once the domain of $L$ is
correctly defined it will be Hermitian and uniqueness of
decomposition will hold and consequently the uniqueness of time
evolution. The mathematical term that we expect to hold is that
$L$ is ``essentially self-adjoint'' namely that it has a unique
Hermitian (i.e. self-adjoint) extension (see \cite{ReedSimon} for
a text book and \cite{HorowitzMarolf,IshibashiHosoya}).

\sbsection{Initial conditions at r=0}

Let us find some necessary boundary conditions at $r=0$ in
response to the ``non-span'' property just discussed.
Since $L$ defined in (\ref{defL}) is singular  at $r=0$
$\del_{tt}$ may be ill-defined (which reflects the singular nature
of the wave equation (\ref{waveeq1})). Since $\phi$ is finite at
$r=0$ we need $\del_{tt} \phi$ to be finite as well
 \footnote{In the rest of this subsection ``finite'' should be
 understood to mean ``finite at $r=0$.''}
and thus we clearly need \be
 L\,  \phi = {\rm finite}
  \label{cond1} \ee
where $\phi$ can be either of the initial conditions $\phi_i,\,
\dot{\phi}_i$.

For RN from the leading behavior for $f_{\rm RN}$
(\ref{fRN-lead}), we get that for a regular and generic $\phi$ the
Laurent expansion of $L \phi$ starts at degree $(-4)$, and thus
eq. (\ref{cond1}) represents 4 conditions (for
degrees $-4 \le {\rm deg} \le -1$). Moreover, if we Taylor expand
$\phi=\sum_{j=0}^\infty \phi^{(j)} \, r^j$ then at order ${\rm
deg}$ the largest $j$ for which $\phi^{(j)}$ appears in the
equation is $j={\rm deg}+6$, and so we have 4 (linear,
homogeneous) conditions for the 6 coefficients $\phi^{(0)}, \dots,
\phi^{(5)}$. These are actually the conditions found in the
paragraph around eq. (\ref{Taylor-phi}), which limit our choice of
initial conditions in the vicinity of $r=0$.

Similarly, for \Schw (\ref{cond1}) represents 3 conditions (for
degrees $-3 \le {\rm deg} \le -1$) among the first 4 Taylor
coefficients which agree with the findings in the paragraph after
eq. (\ref{Taylor-phi}).

However, there are additional requirements: the constraint
(\ref{cond1}) must be compatible with the time evolution according
to the wave equation \be
 L \phi = -\del_{tt} \phi \label{waveeq2} \ee
 namely we must require
 \be \del_{tt} [ L\, \phi] = {\rm finite}
 \ee
 By the wave equation (\ref{waveeq2}) this is equivalent to \be
 L^2\,  \phi = {\rm finite}
  \label{cond2} \ee
This gives 4 additional conditions (in addition those of
(\ref{cond1})) involving the first 12 Taylor coefficients for RN
and 3 additional conditions for the first 8 Taylor coefficients
for Schwarzschild.

\hspace{0.5cm}

Now we may readily generalize, and find \\
{\bf Necessary boundary condition}:   \be L^n\, \phi= {\rm finite}
 ~~ \forall n \ge 1 \label{bc} \ee
 (at $r=0$).

\hspace{0.5cm}

This condition holds both for RN and Schwarzschild and is
explicitly compatible with time evolution. Note that it can also
be interpreted as $\del_{tt}^n\, \phi={\rm finite} ~~ \forall n
\ge 1 $.  It is plausible to us that this condition is not only
necessary but also sufficient for a unique time evolution as well
since showing that $\phi$ has finite time derivatives of any order
at the initial moment of time comes close to providing a
well-defined time evolution, but we shall not attempt to pursue
this point.

As we already mentioned, the boundary conditions (\ref{bc}) are
automatically satisfied for characteristic b.c. (at past null
infinity) and more generally b.c. which vanish at a neighborhood
of $r=0$.

Another perspective is to consider a numerical computer
implementation of this time evolution. One may either have no grid
points at $r=0$ in which case the time evolution is well-defined,
or put a grid point at $r=0$ and find its time increment as a
limit of the time increments of neighboring points. The
singularity will tend to ``expand'' numerical errors, but those
should be possible to tame by decreasing the grid spacing. It
would be interesting to analyze this further and/or to perform
the implementation and determine the behavior at $r=0$.

\subsection{General conditions for wave-regularity}

Having shown that \RN is wave-regular allowing for transmission
across the singularity, we would like to abstract the general
conditions on a spacetime to have such a resolution.

The crucial property of the RN singularity is that the two
solutions $\phiom$ have a unique continuation across $r=0$, while
{\it a priori} there could have been multi-valued functions such
as the log's which appear in the Schwarzschild case. Hence a wave
packet constructed from the $\phiom$ has a prospect of crossing
the singularity in a well-defined manner. This is true not only in
4d RN where the time-separated ODE is regular but also for $d>4$
RN where the ODE becomes regular-singular rather than smooth, but
nevertheless the two characteristic exponents are integral and
there are no $\log$ pieces so that the functions $\phiom$ are
univalued in the vicinity of the singularity. The condition on the
characteristic exponents can be generalized to allow for a
reparameterization of coordinate  $r \to \tilde{r}\sim r^{c_1}$
and a linear redefinition of the field $\phi \to r^{c_2} \,
\phiom$ for some constants $c_1,\, c_2$. In general $\phiom$ can
be series expanded as $\phiom = \tilde{r}^\rho\,
\sum_{k=0}^{\infty}\, \phi^{(k)}\, \tilde{r}^{k\, \Delta}$ where
$\rho=\rho_{1,2}$ is one of the two characteristic exponents and
$\Delta$ is the ``series step-size'' ($\Delta=1$ when the
functions in the differential equation are meromorphic) and
$\phi^{(k)}$ are some constants. The quantity
$(\rho_1-\rho_2)/\Delta$ is invariant under the double
transformation above, and hence there exists a transformation such
that all three $\rho_1,\, \rho_2$ and $\Delta$ are integral and
the function is univalued exactly if the ratio above is rational.

\hspace{0.5cm}

We summarize the above by  \\
 {\bf Necessary general condition}:
the ``eigen-functions'' $\phiom$ should be univalued in the
vicinity of the singularity. This is equivalent to: \begin{itemize}
 \item The equation for $\phiom$ is either regular or regular-singular. If it
is regular-singular we also require the next items:
 \item The difference of
characteristic exponents is commensurate with the ``series
step-size'' (defined in the previous paragraph) \be
 {\rho_2 - \rho_1 \over \Delta} \in \IQ~. \ee
\item There are no $\log$ pieces in the solutions.
\end{itemize}

\hspace{0.5cm}

\section{Summary and discussion}
\label{discussion}

In this paper we saw that physical predictability can be restored
in the neighborhood of the charged (RN) black-hole singularity as
well as negative mass Schwarzschild, once one considers waves
rather than particles. For RN this is done by gluing two
spacetimes over the time-like singularity, thereby adding another
asymptotic region. An observer at infinity in the additional
region views a spacetime with a negative mass. Several
alternatives exist for the total Penrose diagram depending on the
relative size of $M^2$ and $Q^2$. This singularity appears to have
the physical nature of a beam splitter.

For negative-mass Schwarzschild there is a natural ``regularity''
boundary condition at the singularity cutting off spacetime there,
and thus it can be physically interpreted as a perfectly
reflecting mirror.

We subjected this picture to several tests by adding perturbations
in section \ref{perturbations-section}. Considering a field with
both mass and charge we found that such terms are subleading at
the neighborhood of the singularity. Considering added
interactions the field equation becomes non-linear and we were
satisfied in confirming that there is no obstruction (such as
creating a divergence) in extending the linear solutions to next
order. Similarly we then considered the effect of non-linearities
from back-reaction, and found that in some cases a simple argument
protects the regularity of $\phi$ at the next order. Altogether
the results were surprisingly resilient to perturbations.

It would be interesting to perform some additional tests and generalizations:

\begin{itemize}
\item  Study fields with other spin.
Of particular interest are the electromagnetic and gravitational
fields, and also the Dirac $s=1/2$ field. A preliminary
investigation of the electromagnetic and gravitational fields in
d=4 indicates that their polar modes behave just as scalar fields,
i.e. 75\% reflection and 25\% transmission for high energy at
fixed $l$. The axial modes appear to be more subtle, however.

\item  Study other backgrounds such as extreme RN and rotating black-holes.
\end{itemize}

Rotating black-holes deserve a special discussion here. They may
be linked to our work in two different ways. First, the motivation
to the present analysis partly emerges from the desire to explore
the physical phenomena that may take place deep inside realistic
rotating black holes. From this point of view the spherical
charged black hole serves as a toy model for the more complicated,
non-spherical, spinning black hole. In the second link we can
employ the spinning black holes to test the singularity-resolution
approach developed here. In the Kerr-Newman solution the $r=0$
singularity forms a ring rather than a hypersurface. Once the
field is decomposed into spheroidal-harmonic modes, the field
equation for each mode is perfectly regular at $r=0$, hence there
is no doubt about the proper continuation. Now, when the spin
parameter $a$ is taken to vanish, the ring's radius shrinks to
zero, and the spacetime becomes RN. One may therefore {\it define}
the RN extension of the field beyond the $r=0$ singularity to be
the limit $a \to 0$ of the corresponding field in the Kerr-Newman
case. The obvious question is, therefore: does this procedure
yield exactly the same extension as that constructed above? At
least for a scalar field in 4d RN the answer is found to be
positive. It still remains to check whether this is also the case
in the $Q=0$ case. Namely, in the uncharged Kerr case, when $a \to
0$ and the spacetime becomes Schwarzschild, does one recover the
full-reflection b.c. advocated above? This still needs to be
verified.

We would like to mention several other issues at the classical level:

\begin{itemize}
\item  In all of our examples one of the spacetimes has a negative mass
and a globally-naked singularity. Such physical objects raise
problematic issues, and we name only a few: anti-gravity,
acceleration reversed to force, and also inconsistency of their
construction with the Cosmic Censorship conjecture (see
``Perspectives on $r<0$'' in section \ref{ReviewRN}).

It would be interesting to determine whether Nature allows the
actual construction of any spacetime with a {\it wave-regular
timelike singularity}. The mechanism that immediately suggests
itself would be to reconsider a gravitational collapse of a
charged spherical shell (in the positive-mass universe), leading
to a RN black-hole geometry (outside the collapsing object).
However, the instability of the inner horizon raises some doubts
about whether the RN-like $r=0$ singularity will indeed form in
this process.

\item  Implications for {\it Cosmic Censorship}. At the very least it
shakes its rationale since given ``reasonable'' initial conditions
Cosmic Censorship is supposed to ``protect us'' from loss of
predictability, namely of losing unique time evolution, by
forbidding naked singularities, but here we see that such
singularities may not mean the loss of predictability after all.

\item  Since for RN we advocate a picture where there are two spacetimes
which are consistently glued at the singularity
--- and since a macroscopic measuring device cannot cross the singularity
(at best it will bounce back, but it may also be destroyed by
tidal forces, or be ``beam-splitted,'' in the worse case) one may
take {\it two different points of view} on the physics
corresponding to either of the two observers on the two sides of
the singularity. Namely, the story of the resolution of each
singularity will be told in two different versions corresponding
to the two observers (this applies to both types of RN spacetime,
namely $|Q|<|M|$ and $|Q|>|M|$ -- see figures
\ref{PenroseRN},\ref{PenroseRN-ext}).

\end{itemize}

Going beyond classical GR there are open questions as to the quantum gravity
consistency and properties:

\begin{itemize}
\item We must note that our wave equation is outside its {\it domain of
validity} near the singularity due to the presence of high
curvature. However, a preliminary analysis shows that quantum
corrections do not produce essential singularities in the
equations and start altering the field only at a Planck (proper)
distance from the singularity.

\item  Semi-classical quantization and Hawking evaporation (under study).

\item  As one passes to the quantum theory one may
suspect that the pathologies associated with negative-mass spaces
to only grow worse. Therefore, we need to be very cautious in
discussing possible implications for {\it  ``legitimizing''
spacetimes} with such singularities as possible solitons, namely
which spacetimes should be considered to contribute to the path
integral as admissible saddle points. Here we took an open-minded
approach of exploration and we hope that the various interesting
issues which get raised will be studied further.
\end{itemize}

\vspace{0.5cm} \noindent {\bf Acknowledgements}

We would like to thank O. Aharony, J. Bekenstein, M. Berkooz, G.
Gibbons, J. Katz, D. Kazhdan,  N. Itzhaki, D. Kutasov, M. Rozali
and R.M. Wald for discussions. This work is supported in part by
the Israeli Science Foundation. AG is supported in part by the
Israel Academy of Sciences and Humanities -- Centers of Excellence
Program, the German-Israel Bi-National Science Foundation, and the
European RTN network HPRN-CT-2000-00122. BK is supported in part
by the Israeli Science Foundation and by the Binational Science
Foundation BSF-2002160. AS is supported in part by the Horowitz
Foundation.

\appendix

\section{\KS coordinates}
\label{KSapp}

For completeness, we recall here the derivation of the metrics in
the \KS coordinates and the definition of the functions $g(r)$
which are mentioned in the text and are used to implicitly define
$r$ in these coordinates.

First one defines the ``tortoise'' coordinate $r^*$ by \be
 dr^* := {dr \over f} ~. \ee
$r^*$ diverges on horizons but is finite at $r=0$. Then one passes
to light-cone coordinates \bea
 v &:=& t+r^* \non
 u &:=& t-r^*  \eea
and finally in the neighborhood of a horizon with surface gravity
$\kappa$ (for \RN we have $\kappa_\pm = (r_+-r_-)/(2\, r_\pm^2)$
while for \Schw $\kappa=1/(2\, r_0)$)
 the Kruskal-Szekeres coordinates are given by \bea
 V &:=& \pm \exp (\pm \kappa\, v)  \non
 U &:=& \pm \exp (\pm \kappa\, u)  \label{defUV} \eea
and the sign inside the exponent is chosen such that the horizon
is located at $UV=0$, namely it is X-shaped.

In these coordinates the metric is given by \be
 ds^2 = -{f \over \kappa^2\, g}\, dU\, dV + r^2\, d\Omega^2 ~, \ee
 where the function $g(r)$ is defined by \be
 g(r) := \exp (\pm 2\, \kappa\, r^* ) \ee
 and the sign is chosen to be the same as that for $V$ in (\ref{defUV}).
$r$ is not
a coordinate anymore, but rather it is implicitly defined in terms
of the $U,\, V$ coordinates by \be
 -U\, V = g(r) ~. \ee
Note that by construction $g(r)$ has a zero at the horizon and
thus the horizon is explicitly smooth in these coordinates since
in the prefactor of $dU\, dV$ the zero in $f$ gets cancelled
against the zero in $g$.

In \RN there are two horizons at $r_\pm$, and accordingly two
Kruskal-Szekeres-like planes for $(U_\pm,\, V_\pm)$, and two
functions $g_\pm (r)$. The ``upper plus'' quadrant ($U_+,V_+>0$)
is glued to the ``lower minus'' quadrant ($U_-,V_-<0$) according
to \bea
 V_+^{1/\kappa_+} &=& (-V_- )^{1/\kappa_-} \non
 U_+^{1/\kappa_+} &=& (-U_- )^{1/\kappa_-}  \eea
and similarly one may continue and glue the upper minus quadrant
to another copy of the plus plane, creating the maximal analytic
extension of \RN which consists of an infinite chain of
alternating plus and minus planes. $g_\pm(r)$ are given by (see
figure \ref{gRNfigure}) \bea
 g_+(r) &=& \({r \over r_+}-1 \) \, \({r \over r_-}-1 \)^{-{\kappa_+ \over
\kappa_-}} \, \exp (2\, \kappa_+\, r) \non
 g_-(r) &=& \(1-{r \over r_-}\) \, \(1-{r \over r_+}\)^{-{\kappa_- \over
\kappa_+}} \, \exp
 (-2\, \kappa_-\, r) \eea
While for Schwarzschild we have (see figure \ref{gSchw-figure})
\be
 g_{\rm schw}(r) = \({r \over r_0}-1 \)\, \exp(r/r_0) ~. \ee
Finally, in order to get the Penrose diagram (see figures
\ref{PenroseRN},\ref{PenroseSchw},\ref{PenroseRN-ext}) one
customarily takes the conformal transformation \bea
 U_P = {\rm tg}^{-1} (U) \non
 V_P = {\rm tg}^{-1} (V) \eea
although any other transformation which maps the real line to an
interval is admissible.

\section{Regular singularities of ordinary differential equations}
\label{reg-sing-app}

Let us briefly recall the definitions.
A linear second order differential equation \be
 [a(r)\, \del_{rr} + b(r)\, \del_r + c(r)]\, \phi=0 \ee
is regular-singular at a point $r_0$, which we will assume without
loss of generality to be $r_0=0$, if after normalization such
 that $a(0)=1$ there are singularities at $r=0$ in $b(r)$ or
$c(r)$ of limited type: $b(r)$ may have at most a first order
pole, and $c(r)$ at most a second order pole. This guarantees that
the solutions will have at most poles or branch cuts at $r=0$ but
not an essential singularity. More specifically the two solutions
have the leading behavior \be
 \phi_{1,2} \sim r^{\rho_{1,2}} ~, \ee
 where $\rho_{1,2}$ are called the characteristic exponents, and they are
the solutions to the quadratic equation \be
 a(r_0) \rho\, (\rho-1) + b(r_0)\, r  \rho + c(r_0)\, r^2  =0 \ee
 gotten from substituting the leading
behavior above into the equation. When $\rho_1=\rho_2=\rho$ the
leading behavior is $\phi_1=r^\rho, ~ \phi_2=r^\rho \, \log(r)$.
Finally, one can expand the solutions into a series $\phi = r^\rho
\sum_{j=0}^{\infty}\, \phi^{(j)}\, r^j$ where $\phi^{(j)}$ are
some constants, except that in the case when $\rho_2-\rho_1$ is a
positive integer $\phi_1$ may contain also a piece proportional to
$\log(r)\, \phi_2$ and thus contains a log.

\section{Basis functions for the characteristic formulation}
\label{char-appendix}

Here we define the basis functions $\phi _{(\omega _{v},0)}^{*},\,
\phi _{(0,\omega _{u})}^{*}$ for the characteristic formulation
(see subsection \ref{char-subsection}). Since the radial equation
is a second-order ODE, the radial functions $\phi _{\omega
}^{*}(r^*)$ for a given $\omega $ form a two-parameter family.
Owing to the asymptotic behavior of the radial functions, Eq.
(\ref{gen-asymptotic}), we may choose two basis functions $\phi
_{\omega (1,0)}^{*}(r^*)$ and $\phi _{\omega (0,1)}^{*}(r^*)$,
defined as follows: $\phi _{\omega (1,0)}^{*}(r^*)$ is the radial
function which at $r^*\to +\infty $ (the inner horizon) has the
asymptotic form
\[
\phi _{\omega (1,0)}^{*}\cong T(\omega ){\rm exp}(-i\,\omega
\,r^*)
\]
and at $r^*\to -\infty $ (the negative-$r$ asymptotically-flat
region) has the asymptotic form
\[
\phi _{\omega (1,0)}^{*}\cong {\rm exp}(-i\,\omega \,r^*)+R(\omega
){\rm exp} (+i\,\omega \,r^*),
\]
where $T(\omega )$ and $R(\omega )$ are two unconstrained
coefficients (these coefficients turn out to be the transmission
and reflection coefficient; see section
\ref{cross-section-section}). Similarly, $\phi_{\omega
(0,1)}^{*}(r^*)$ is the radial function which at $r^* \to -\infty
$ has the asymptotic form
\[
\phi _{\omega (0,1)}^{*}\cong T^{\prime }(\omega ){\rm
exp}(+i\,\omega \,r^*)
\]
and at $r^*\to +\infty $ has the asymptotic form
\[
\phi _{\omega (0,1)}^{*}\cong {\rm exp}(+i\,\omega
\,r^*)+R^{\prime }(\omega ) {\rm exp}(-i\,\omega \,r^*),
\]
where again $T^{\prime }(\omega )$ and $R^{\prime }(\omega )$ are
two unconstrained coefficients. Now, the two desired solutions
$\phi _{(\omega _{v},0)}^{*}$ and $\phi _{(0,\omega _{u})}^{*}$
are simply given by
\begin{eqnarray*}
\phi _{(\omega _{v},0)}^{*}(r^*,t) &=&\phi _{\omega
_{v}(1,0)}^{*}(r^*){\rm exp}(i\,\omega _{v}\,t)\,,\, \\
\phi _{(0,\omega _{u})}^{*}(r^*,t) &=&\phi _{\omega
_{u}(0,1)}^{*}(r^*){\rm exp}(i\,\omega _{u}\,t).
\end{eqnarray*}
As one can easily verify, these two solutions take the asymptotic
forms
\[
\phi _{(\omega _{v},0)}^{*}\cong T(\omega _{v}){\rm exp}(i\,\omega
_{v}\,v)\qquad \qquad (r^*\to +\infty ),
\]
\[
\phi _{(\omega _{v},0)}^{*}\cong {\rm exp}(i\,\omega
_{v}\,v)+R(\omega ){\rm exp}(i\,\omega _{v}\,u)\qquad \qquad
(r^*\to -\infty ),
\]
and
\[
\phi _{(0,\omega _{u})}^{*}\cong T^{\prime }(\omega _{u}){\rm
exp}(i\,\omega _{u}\,u)\qquad \qquad (r^*\to -\infty ),
\]
\[
\phi _{(0,\omega _{u})}^{*}\cong {\rm exp}(i\,\omega
_{u}\,u)+R^{\prime }(\omega ){\rm exp}(i\,\omega _{u}\,v)\qquad
\qquad (r^*\to +\infty ),
\]
in agreement with the above definitions of $\phi _{(\omega
_{v},0)}^{*}$ and $\phi _{(0,\omega _{u})}^{*}$.

\section{Bessel functions}
\label{BesselApp}

Let us assemble a few useful properties of the Bessel functions. A
Bessel function of order $n$ satisfies the equation
$[\del_{xx}+(1/x)\, \del_x + 1-n^2/x^2 ]\,J_n\, =0$. It will be
more convenient for us to use the equivalent definition \be
  \br \del{xx} + \(1- {n^2-1/4 \over x^2} \)  \kt \, \sqrt{x} J_n(x)
 =0 ~. \ee
For small $x$ \be J_n(x) = x^n\, A^{J}_n(x^2) ~,
\label{BesselSing} \ee
 where $A^{J}_n(x^2)$ are certain functions of $x^2$ analytic at $x^2=0$.
The asymptotic behavior is better described by \be
 Y_n=(J_n\, \cos (n\, \pi) - J_{-n})/\sin(n\, \pi) \ee
 and one has \be
 J_n \pm i\, Y_n \simeq \sqrt{2/(\pi\, x)}~ \exp \left[ \pm i(x-n\,
 \pi/2-\pi/4) \right]
 \label{BesselAsymp}\ee
 as $x \to +\infty$.


\end{document}